\documentclass[twocolumn]{aastex63}
\usepackage{graphicx}	
\usepackage{amssymb}
\usepackage{array}
\usepackage{verbatim}
\usepackage{epstopdf}
\usepackage{epsfig}
\usepackage[title]{appendix}

\newcommand{\mum}{$\mu$m}
\newcommand{\eg}{e.\,g.,}
\newcommand{\ruv}{$r(u,v)$}
\newcommand{\ruvmin}{$r$($u$,$v$)$_{\rm min}$}

\newcommand{\mm}{$\pm$}
\newcommand{\xx}{$\times$}
\newcommand{\rsun}{~R$_{\odot}$~}
\newcommand{\msun}{~$M_{\sun}$}
\newcommand{\lsun}{~$L_{\sun}$~}
\newcommand{\Macc}{$\dot{M}_{\rm acc}$}

\newcommand{\msunyr}{~$M_{\sun}$~yr$^{-1}$}
\newcommand{\mJybeam}{~mJy beam$^{-1}$}
\newcommand{\uJybeam}{~$\mu$Jy beam$^{-1}$}
\newcommand{\mdisk}{~$M_{\mathrm{disk}}$}
\newcommand{\rdisk}{~$R_{\mathrm{disk}}$}

\shorttitle{Modeling disk around massive YSO}
\shortauthors{A\~nez-L\'opez et al.}
\begin{document}

\title{Modeling the accretion disk around the high-mass protostar GGD 27-MM1}

\author{N. A\~nez-L\'opez}
\affil{Institut de Ci\`encies de l'Espai (ICE, CSIC), Can Magrans s/n, E-08193 Cerdanyola del Vall\`es, Catalonia}
\affiliation{Institut d'Estudis Espacials de Catalunya (IEEC), E-08034 Barcelona, Catalonia}

\author{M. Osorio}
\affiliation{Instituto de Astrof\'{\i}sica de Andaluc\'{\i}a (IAA, CSIC), Glorieta de la Astronom\'{\i}a s/n, E-18008 Granada, Spain}

\author[0000-0002-2189-6278]{G. Busquet}
\affil{Institut de Ci\`encies de l'Espai (ICE, CSIC), Can Magrans s/n, E-08193 Cerdanyola del Vall\`es, Catalonia}
\affiliation{Institut d'Estudis Espacials de Catalunya (IEEC), E-08034 Barcelona, Catalonia}

\author[0000-0002-3829-5591]{J. M. Girart}
\affil{Institut de Ci\`encies de l'Espai (ICE, CSIC), Can Magrans s/n, E-08193 Cerdanyola del Vall\`es, Catalonia}
\affiliation{Institut d'Estudis Espacials de Catalunya (IEEC), E-08034 Barcelona, Catalonia}

\author{E. Mac\'ias}
\affil{Department of Astronomy, Boston University, 725 Commonwealth Avenue, Boston, MA 02215, USA}
\affil{Joint ALMA Observatory, Alonso de C\'ordova 3107, Vitacura, Santiago 763-0355, Chile}
\affil{European Southern Observatory, Alonso de C\'ordova 3107, Vitacura, Santiago 763-0355, Chile}

\author{C. Carrasco-Gonz\'alez}
\affil{Instituto de Radioastronom\'ia y Astrof\'isica (IRyA), UNAM, Apdo. Postal 72-3 (Xangari), Morelia, Michoac\'an 58089, M\'exico}

\author[0000-0003-4576-0436]{S. Curiel}
\affil{Instituto de Astronom\'ia, Universidad Nacional Aut\'onoma de M\'exico (UNAM), Apartado Postal 70-264, DF 04510 M\'exico}

\author{R. Estalella}
\affil{Departament de F\'isica Qu\`antica i Astrof\'isica, Institut de Ci\`encies del Cosmos (ICC), Universitat de Barcelona (IEEC-UB), Mart\'i i Franqu\`es 1, 08028, Catalonia}

\author{M. Fern\'andez-L\'opez}
\affil{Instituto Argentino de Radioastronom\'ia (CCT-La Plata, CONICET; CICPBA), C.C. No. 5, 1894, Villa Elisa, Buenos Aires, Argentina}

\author[0000-0003-1480-4643]{R. Galv\'an-Madrid}
\affil{Instituto de Radioastronom\'ia y Astrof\'isica (IRyA), UNAM, Apdo. Postal 72-3 (Xangari), Morelia, Michoac\'an 58089, M\'exico}

\author{J. Kwon}
\affil{The University of Tokyo, Hongo 7-3-1, Bunkyo, Tokyo 113-0033, Japan}

\author{J.M. Torrelles}
\affil{Institut de Ci\`encies de l'Espai (ICE, CSIC), Can Magrans s/n, E-08193 Cerdanyola del Vall\`es, Catalonia}
\affiliation{Institut d'Estudis Espacials de Catalunya (IEEC), E-08034 Barcelona, Catalonia}

\begin{abstract}

Recent high-angular resolution ($\simeq 40$ mas) ALMA observations at 1.14~mm resolve a compact ($R \simeq 200$ au) flattened dust structure perpendicular to the HH~80--81 jet emanating from the GGD~27-MM1 high-mass protostar, making it a robust candidate for a true accretion disk. The jet/disk system (HH~80--81/GGD~27-MM1) resembles those found in association with low- and intermediate-mass protostars. We present radiative transfer models that fit the 1.14~mm ALMA dust image of this disk which allow us to obtain its physical parameters and predict its density and temperature structure. Our results indicate that this accretion disk is compact (\rdisk $\simeq$ 170 au) and massive ($\simeq5$ \msun), about 20\% of the stellar mass of $\simeq 20$\msun. 
We estimate the total dynamical mass of the star-disk system from the molecular line emission finding a range between 21 and 30\msun, which is consistent with our model.
We fit the density and temperature structures found by our model with power law functions. 
These results suggest that accretion disks around massive stars are more massive and hotter than their low-mass siblings, but they still are quite stable. 
We also compare the temperature distribution in the GGD~27--MM1 disk with that found in low- and intermediate-mass stars and discuss possible implications on the water snow line.
We have also carried out a study of the distance based on Gaia DR2 data and the population of young stellar objects (YSOs) in this region, and from the extinction maps. We conclude that the source distance is within 1.2 and 1.4~kpc, closer than what was derived in previous studies (1.7~kpc). 
\end{abstract}

\keywords{stars: formation -- ISM: individual objects (GGD~27, HH~80--81, IRAS\,18162-2048) -- stars: massive -- protoplanetary disks}

%%%%%%%%%%%%%%%%%%%%%%%%%%%%%%%%%%%%%%%%%%%%%%%%%%%%%%%%%%%%%%%%%%%%%%%%%
\section{Introduction} \label{sec:intro}

Understanding how high-mass stars form and evolve is one of the hot topics in astrophysics, due to the strong impact that these objects have in the life of a galaxy. However, the study of high-mass protostars is difficult due to their fast evolution ($\sim10^5$~yr) to the main sequence, their large distances and high obscuration. 

It is well known that low-mass stars are formed through an accretion disk that transports gas and dust from the envelope to the protostar and through a jet that removes the excess of angular momentum \citep*{ShuAdamsLizano87,McKee&Ostriker2007}.
Disks around nearby solar-type stars have been studied to great extent and detail \citep[e.g.][]{Williams&Cieza2011, Testi2014, HartmannHerczegCalvet2016, Andrews18}, but at the moment the number of disks studies of more distant and massive stars is still comparatively very small. 

Accretion disks around massive stars is a plausible mechanism that can alleviate the radiation pressure problem, hence allowing an accretion flow to continue once photo-ionization has started \citep[e.g.][]{Tan2014, Klassen2016, Kuiper2018}. However, their physical properties are still uncertain \citep[see \eg][for a recent review]{Beltran&deWit2016}. Flattened, disk-like structures have been observed in a few massive young stars
\citep{Beltran2011, Beltran2014, Sanchez-Monge2014, Johnston2015, Sanna2018, Zapata2019}. 
However, these structures are large ($\sim$ 1000--10000 au) and  have masses considerably larger than the central protostar and 
it is difficult to envisage them as real accretion disks. 
Therefore higher angular resolution and more sensitivity observations are required to better characterize the physical parameters and the role of these rotating structures \citep[e.g.][]{Maud2019, Ginsburg2019}.
Three of the best examples in the literature of massive protostars associated with a clearly defined jet and a compact (few hundred au) angularly resolved dusty disk candidate are: Cepheus A HW2 \citep{Patel2005}, GGD~27-MM1 \citep{Girart2018april}, and G11.92-061 MM1a \citep{Ilee2018}. 

The HH~80--81 objects (at a distance of 1.4 kpc; see Appendix) are associated with a spectacular ($\sim10$~pc long) highly collimated radio-jet \citep{Marti1993,Heathcote98, Masque2015}, which is powered by a massive early B-type protostar IRAS\,18162$-$2048 \citep[GGD~27-MM1,][]{Fernandez-Lopez2011, Girart2017}. This protostellar radio-jet is the first one where polarized emission due to relativistic electrons has been detected, showing the presence of a magnetic field aligned with the jet \citep{CarrascoGonzalez2010,Rodriguez-Kamenetzky17}. 
Very Large Array (VLA) continuum observations at 7 mm reveal a cross-shaped morphology which was interpreted as two overlapping structures that could correspond to the radio-jet and a disk of $\sim$200 au of radius \citep{CarrascoGonzalez2012}, in agreement
 
with the upper limit imposed by the 1.3~mm continuum dust emission observations \citep{Fernandez-Lopez2011}. The size of the putative disk coincides with theoretical predictions of the centrifugal radius based on the Spectral Energy Distribution (SED) fitting of some high-mass protostar regions \citep*{deBuizerOsorioCalvet2005}.

\begin{figure}[!hbtp]
\centering
\includegraphics[width=0.98\columnwidth]{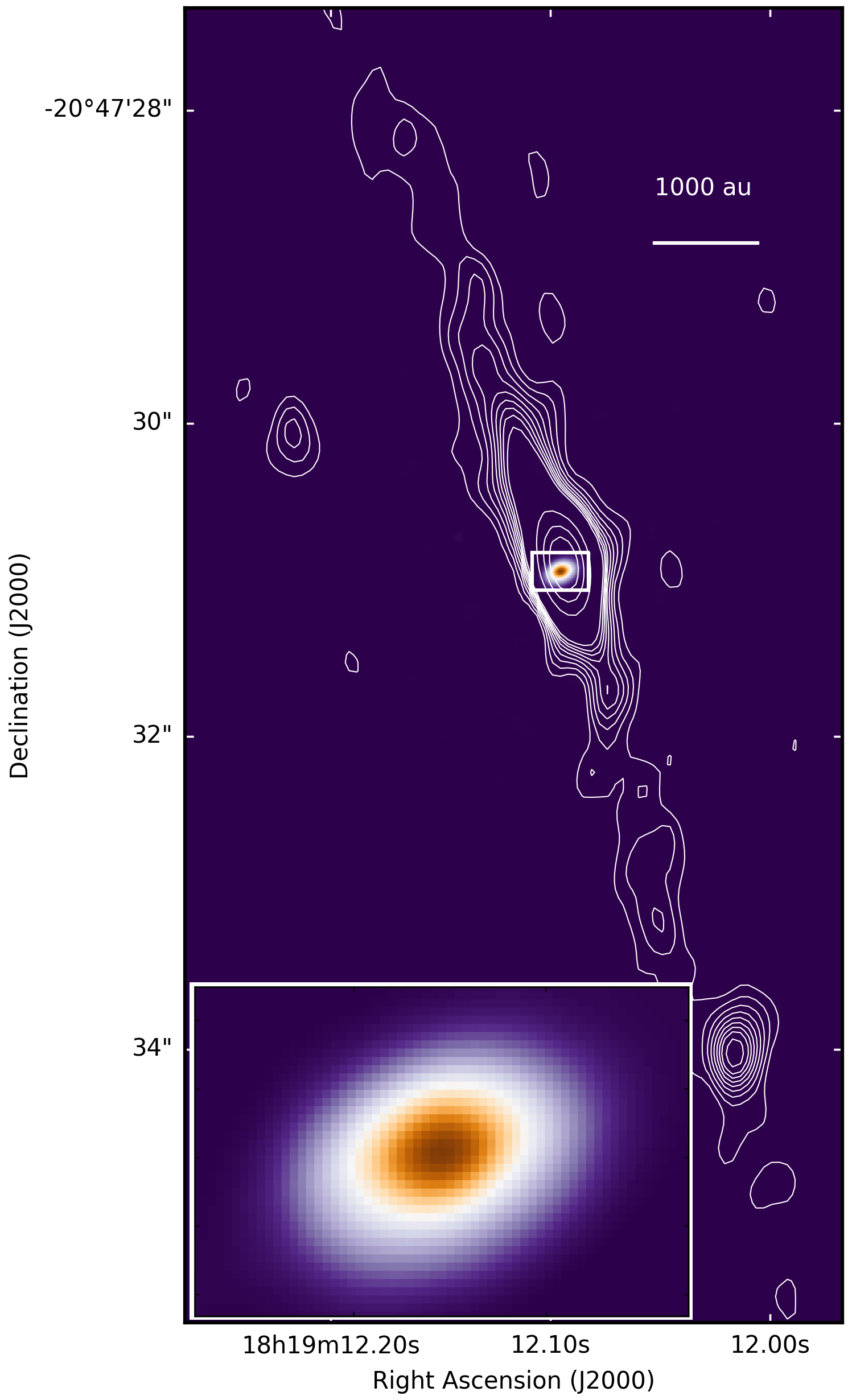}
\caption{\label{fig:disk+jet} Disk and jet system of the massive star GGD~27 -MM1. The color image shows the dust continuum emission of the disk observed with ALMA at 1.14~mm with $\sim$40 mas angular resolution ($\sim$56 au) \citep[][see also our Fig. \ref{fig:Real-uv}]{Girart2018april}.
In contours it is shown the VLA image at 3.6 cm of the radio-jet observed with an angular resolution of $\sim$0.4 arcsec \citep{CarrascoGonzalez2012}.
} 
\end{figure}

In addition, observed velocity gradients in the molecular gas perpendicular to the HH80-81 radio jet have been interpreted as rotating motions \citep{Fernandez-Lopez2011July, CarrascoGonzalez2012, Girart2017}. 
Definitive evidence of a compact disk around IRAS\, 18162-2048 comes from ALMA observations at 1.14~mm with an angular resolution of $\sim$ 40 mas ($\sim$ 56 au), which reveal a compact dust disk clearly perpendicular to the radio-jet \citep[][see Fig.~\ref{fig:disk+jet}]{Girart2018april}.

Due to the similarities with disk-protostar-jet systems in low- and intermediate-mass protostars, in this paper we analyze the GGD~27--MM1 disk by applying models that have successfully explained disks around low-mass stars.
The main goal is to investigate if the assumptions that are usually adopted for disks around low-mass stars can be roughly extrapolated to the case of massive stars. For example, disks around low-mass stars are 
much less massive than the central protostar and therefore they are usually gravitationally stable. In the high-mass case it is not clear if disks are stable \citep{Maud2019} or unstable \citep{Motogi2019,Zapata2019}.

The paper layout is as follows. In Section 2 we present the ALMA observations of the disk around GGD~27--MM1, describing the disk model in Section 3. In Section 4 we present the main observational properties of the region. In Section 5 we explain the fit procedure and Sections 6 and 7 correspond to results and discussion, respectively. We list the main conclusions in Section 8 and present the new distance estimation in the Appendix.

%%%%%%%%%%%%%%%%%%%%%%%%%%%%%%%%%%%%%%%%%%%%%%%%%%%%%%%%%%%%%%%%%%%%%%%%%
\section{ALMA Observations \label{sec:observation}}
%%%%%%%%%%%%%%%%%%%%%%%%%%%%%%%%%%%%%%%%%%%%%%%%%%%%%%%%%%%%%%%%%%%%%%%%%
In this work we use ALMA continuum observations at 1.14~mm (263.0 GHz; project ALMA\#2015.1.00480.S). The Band 6 receiver with the correlator set in continuum mode (time division mode, TDM) covering the 253.0-257.0 GHz and 269.0-273.0 GHZ frequency ranges was used. The observations were carried out with 37 antennas in the c36-8/7 configuration, which provided baselines between 82~m and 11.05~km (13 to 5400~k$\lambda$). 
The Stokes $I$ image toward GGD~27 -MM1 was generated using the resulting visibilities after the subtraction of the compact source (see Section \ref{sec:compact}). This was done using the CASA task {\it tclean} with a value of 0.5 for the robust Briggs weighting parameter. Because of a lack of visibilities between 150 and 300 k$\lambda$ and in order to filter extended emission coming from the envelope, we used visibilities from baselines larger than 300 k$\lambda$ \citep[calibration of the data is described in][]{Girart2018april}. The resulting synthesized beam has a full width at half maximum (FWHM) of 45.0 mas $\times$ 38.3 mas (PA = -62.4$\arcdeg$). The Stokes $I$ rms noise is 60 \uJybeam. 
The ALMA\#2015.1.00480.S project had also a science goal at Band 7 to observe several molecular lines in the 298-302 and 310-313~GHz frequency ranges (c40-6 configuration). Here we analyze the position-velocity diagrams of two of the brightest lines detected that better trace the disk kinematics, SO$_2$ 9$_{2,8}$--8$_{1,7}$ and 19$_{3,17}$--19$_{2,18}$ transitions.
For these two molecular lines, the calibrated data were self-calibrated from the continuum by using all the available channels in the four observed spectral windows, except for the channels with the brightest line emission (e.g., H$_2$CO 4$_{2,3}$-3$_{1,2}$). Velocity channel maps were obtained (after continuum subtraction) using {\it tclean} with natural weighting, which yielded a FWHM synthesized beam of $0.21''\times0.16''$ (PA = $-87.1\arcdeg$). The channel width was $\simeq 0.98$~km~s$^{-1}$. The rms noise achieved was 1.3~mJy~beam$^{-1}$ per channel.
The other molecular lines detected will be presented in a forthcoming paper (Fern\'andez-L\'opez et al. in prep.).

%%%%%%%%%%%%%%%%%%%%%%%%%%%%%%%%%%%%%%%%%%%%%%%%%%%%%%%%%%%%%%%%%%%%%%%%%%
\subsection{The compact (few au) source at the disk center\label{sec:compact}}
\begin{figure}[tbp]
    \centering
    \includegraphics[width=0.99\columnwidth]{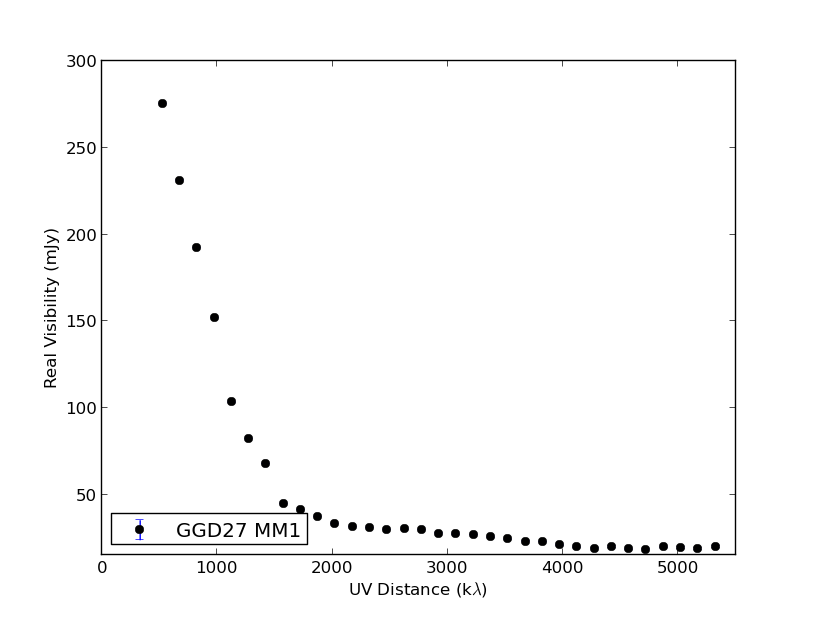}
    \caption{Annular average of the real part of the 1.14 mm ALMA visibilities, centered in GGD~27 MM1, as a function of the $uv$ distance where the error bars are smaller than symbols.}
    \label{fig:ALMadisk-uv}
\end{figure}

\begin{figure}[tbp]
    \centering
    \includegraphics[width=0.95\columnwidth]{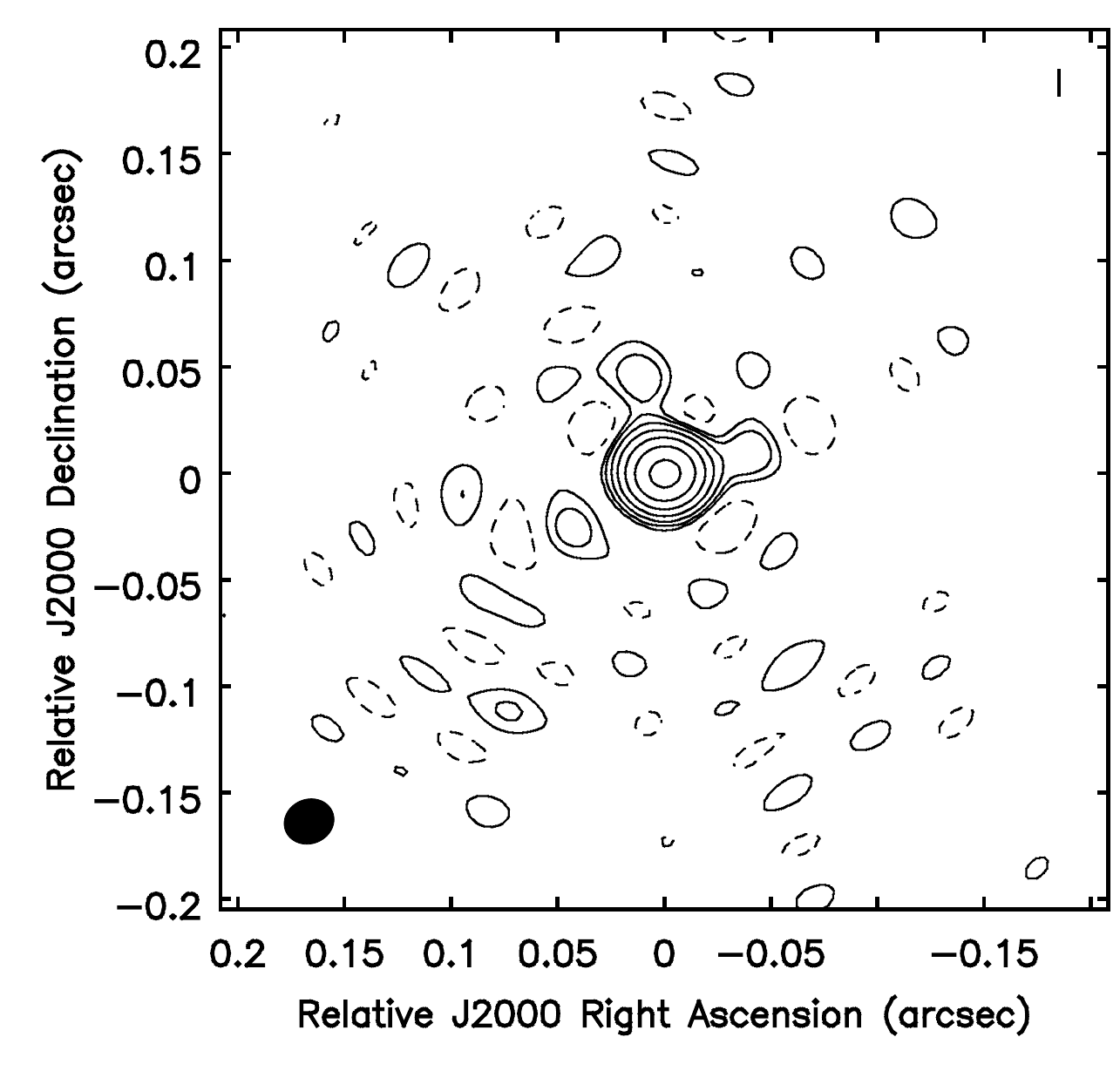}
    \caption{1.14~mm ALMA image of GGD~27 MM1 obtained using only visibilities from baselines larger than $\ge$4000~k$\lambda$. Contour levels start at 4--$\sigma$, where $\sigma$ is 0.06~mJy~beam$^{-1}$, and then increase by a factor two at each contour. The emission shows a compact source with a brightness temperature $\sim$10$^4$K, likely associated with free-free emission (see Section \ref{sec:ionizedcomponent}).}
    \label{fig:ALMcompact}
\end{figure}

%%%%%%%%%%%%%%%%%%%%%%%%%%%%%%%%%%%%%%%%%%%%%%%%%%%%%%%%%%%%%%%%%%%%%%%%%
Figure~\ref{fig:ALMadisk-uv} shows the real part of the observed visibilities as a function of the $uv$ distance from the phase center (disk center).
The flux density decreases steeply with increasing visibility radius, \ruv, for the shortest baselines (\ruv$\la 2000$~k$\lambda$). At larger radii the flux density decreases more smoothly up to \ruv $\simeq$4000~k$\lambda$. At this point the flux density remains roughly constant with the visibility radius. This suggests that the emission from the longest baselines may be dominated by a very compact object. To check this possibility, we generated several images using only visibilities with a minimum visibility radius, \ruvmin, between 3500 and 4750~k$\lambda$ and with a robust weighting of 1. The values of the minimum visibility radius used and the resulting synthesized beam are given in Table~\ref{tab:compact}. 
Maps including only baselines longer than 4000~k$\lambda$ are devoid of artifacts due to the severe missing flux density from the disk. 
The compact source appears in maps with long visibilities radii (see Fig.~\ref{fig:ALMcompact}).

A two dimensional Gaussian fit was performed to the different images obtained with different baseline ranges. The flux density and the deconvolved size obtained from the Gaussian fit are listed in Table~\ref{tab:compact}. The images with visibilities radii $\geq$ 4000~k$\lambda$ show an unresolved source with a flux density of $\sim$19~mJy at the center of the disk. 
We also performed a Gaussian fit to the visibilities using the same range of visibilities as before. The flux density and size of these fits are also shown in Table~\ref{tab:compact}. 
The fits to the visibilities with \ruvmin $>$4000 and $>$4250~k$\lambda$ reveal that the emission arises from a very compact region with a radius of $\sim 4$~mas ($\sim$5.6~au). 
Given the relatively short range of visibilities used in this fit, further very high angular resolution observations ($\sim 10$~mas) are needed to better constrain its size. In any case, these values imply that the brightness temperature of this compact source is probably $\sim 10^4$~K. 

The origin of the compact source cannot be due to the thermal emission of dust grains but rather to ionized gas (see Section \ref{sec:ionizedcomponent}). Therefore, this compact source was removed from the visibilities to obtain the map presented here, tracing only the dust emission from the accretion disk (Fig. \ref{fig:Real-uv}).
%%%%%%%%%%%%%%%%%%%%%%%FIGURES%%%%%%%%%%%%%%%%%%%%%%%%%%%%%%%%%%%%%%%%%

%%%%%%%%%%%%%%%%%%%%TABLES%%%%%%%%%%%%%%%%%%%%%%%%%%%%%%%%%%%%%%%%%%%
\begin{table*}
\caption{Gaussian fit to the Compact Source \label{tab:compact}}
\begin{center}
\begin{tabular}{c c c c c c r}
\hline
\hline
& 
\multicolumn{3}{c}{Image plane} &&
\multicolumn{2}{c}{Visibility plane} \\
\cline{2-4}
\cline{6-7}
\multicolumn{1}{c}{ \ruvmin } 
& Synthesized Beam & Flux density & Deconvolved Size && Flux density & 
\multicolumn{1}{c}{Size} \\
\colhead{(k$\lambda$)} & 
\colhead{(mas\xx mas,\arcdeg)} & 
\colhead{(mJy)} &
\colhead{(mas\xx mas,\arcdeg)} &&
\colhead{(mJy)} & 
\multicolumn{1}{c}{(mas)} \\
\hline		
3500 & 26.7\xx22.4, -63 & 28.5\mm0.2& 16\mm1$\times$7\mm1, 91\mm5 && 29.5\mm0.1 & 13.6\mm0.1 \\
4000 & 23.8\xx20.9, -68 & 21.0\mm0.2& unresolved && 20.8\mm0.1 & 4.7\mm0.5 \\
4250 & 21.9\xx20.9, -59& 20.1\mm0.2 & unresolved && 19.9\mm0.3 & 3.5\mm1.0 \\
4500 & 21.6\xx19.8, 77 & 19.7\mm0.3 & unresolved && 19.3\mm0.4 & failed \tablenotemark{$a$} \\
4750 & 21.8\xx18.3, 67 & 19.1\mm0.4 & unresolved && 19.5\mm0.6 & failed \tablenotemark{$a$}\\
\hline		
\end{tabular}
\tablenotetext{$a$}{Algorithm did not found solution probably due to the small range of visibilities.}
\end{center}
\end{table*}

%%%%%%%%%%%%%%%%%%%%%%%%%%%%%%%%%%%%%%%%%%%%%%%%%%%%%%%%%%%%%%%%%%%%%%%%%
\section{Disk model} \label{sec:model}
%%%%%%%%%%%%%%%%%%%%%%%%%%%%%%%%%%%%%%%%%%%%%%%%%%%%%%%%%%%%%%%%%%%%%%%%%
The disk was modeled using the irradiated $\alpha$-accretion disk models created by \cite{D'Alessio1998, D'Alessio1999, D'Alessio2001, D'Alessio2006}, which have been successfully used and further developed to model disks around low- and intermediate-mass stars \citep[e.g.;][]{McClure_2013,Osorio2014,Osorio2016,Macias2018}.

The models describe disks around stars with parameters typical of classical T Tauris; that is, an irradiated flared disk with two population of grains.
These two populations aim at emulating the dust growth and vertical settling predicted by dust evolution models \citep{DullemondandDominik2004}.
The code computes the vertical and the radial structure of the disk using the $\alpha$-viscosity prescription and enforcing vertical hydrostatic equilibrium. 
In the past, these models have already been used to reproduce the spectral energy distribution (SED) of a possible disk around the high-mass protostar AFGL~2591-VLA~3 \citep{Trinidad2003}. Furthermore, the $\alpha$-viscosity prescription has been also used to model quasi-steady selfgravitating disks around massive protostars under certain conditions ($H/R$ $\leq$ 0.1 and \mdisk/$M_*$ $<$ 0.5, with $H$ the disk scale height, and $R$ the radius of the disk; \citealt{Forgan_2016}).

The model allows to set two populations of grains, with a power-law size distribution $n(a) \propto a^{-3.5}$, where $a$ is the grain radius. 
Regarding the degree of settling, we used the epsilon parameter $\epsilon = \zeta_{\mathrm{small}}/ \zeta_{\mathrm{std}}$, where  $\zeta_{\mathrm{small}}$ and $\zeta_{\mathrm{std}}$ are the dust-to-gas mass ratio of the small grains and the initial standard value respectively \citep[see][]{D'Alessio2006}. 
The relative abundances of the different dust components were adopted with a dust-to-gas ratio of 0.0065 corresponding to the measured abundances of silicates and graphites in the ISM \citep{Draine&Lee1984, D'Alessio2006, Osorio2014}. The remaining ratio up to the commonly used value of 0.01 would be water ice and other ices. These ices sould be sublimated at the high temperatures expected in the disk, so they should have a negligible contribution in the model (see Section \ref{subsec:snow-line}.)

Other parameter related with the settling is $Z_{\mathrm{big}}$ that locates, as a function of the scale height ($H$), the border between both populations of grains, which was fixed as $Z_{\mathrm{big}}$=0.1$H$. 

The main heating sources are the stellar irradiation and the viscous dissipation, which is parameterized through $\alpha$ \citep{Shakura&Sunyaev_1973} and it is assumed to be constant over the disk. The viscosity effective coefficient is defined as $\nu_t~=~\alpha c_{\mathrm{s}} H$,  where $c_{\mathrm{s}}$ is the local sound speed, and $H$ is the hydrostatic scale height of the gas: 
\begin{equation}
    \label{eq:scaleheight}
    \frac{H}{R}=\frac{c_{\mathrm{s}} \left(T_{\mathrm{c}}\right)}{R \Omega (R)}=\left[\frac{k T_{\mathrm{c}} R}{G M_{\mathrm{tot}} \mu m_{\mathrm{H}}}\right]^{1/2},
\end{equation}
where $k$ is the Boltzmann constant, $T_{\mathrm{c}}$ is the disk mid-plane temperature, $G$ is the gravitational constant, 
$M_{\mathrm{tot}}$ is the total mass ($M_*$ + $M_{\mathrm{disk}}$) at every radius, $\mu$=2.33 is the mean molecular weight and $m_{\mathrm{H}}$ is the hydrogen mass. Besides, the model considers accretion luminosity as part of the irradiation of the disk.

The temperature and density structure are calculated self-consistently once the stellar parameters (radius $R_*$, mass $M_*$, and temperature $T_*$), the dust content (abundances, distribution of grain sizes),  viscosity ($\alpha$), and disk mass accretion rate (\Macc) are set.
The dust opacity includes absorption and isotropic self-scattering.
In an $\alpha$-accretion disk model the mass surface density is $\Sigma$=\Macc$\Omega$/3$\pi\alpha$$c_{\mathrm{s}}$($T_{\mathrm{c}}$)$^2$.
The remaining parameters to describe the disk model are: the disk radius, $R_{\rm disk}$, and the inclination angle of the disk $i$.

The disk is considered to be steady, axisymmetric, and geometrically thin. Its self-gravity is neglected compared to the stellar gravity, and it is assumed to be in Keplerian rotation and in hydrostatic equilibrium in the vertical direction. The model assumes that dust and gas are well mixed and thermally coupled; thus, a unique temperature as a function of position in the disk is calculated.

%%%%%%%%%%%%%%%%%%%%%%%%%%%%%%%%%%%%%%%%%%%%%%%%%%%%%%%%%%%%%%%%%%%%%%%%%
\section{The GGD~27-MM1 disk-jet system \label{sec:HH80--81}}
%%%%%%%%%%%%%%%%%%%%%%%%%%%%%%%%%%%%%%%%%%%%%%%%%%%%%%%%%%%%%%%%%%%%%%%%%

\begin{figure}
    \centering
    \includegraphics[width=0.98\columnwidth]{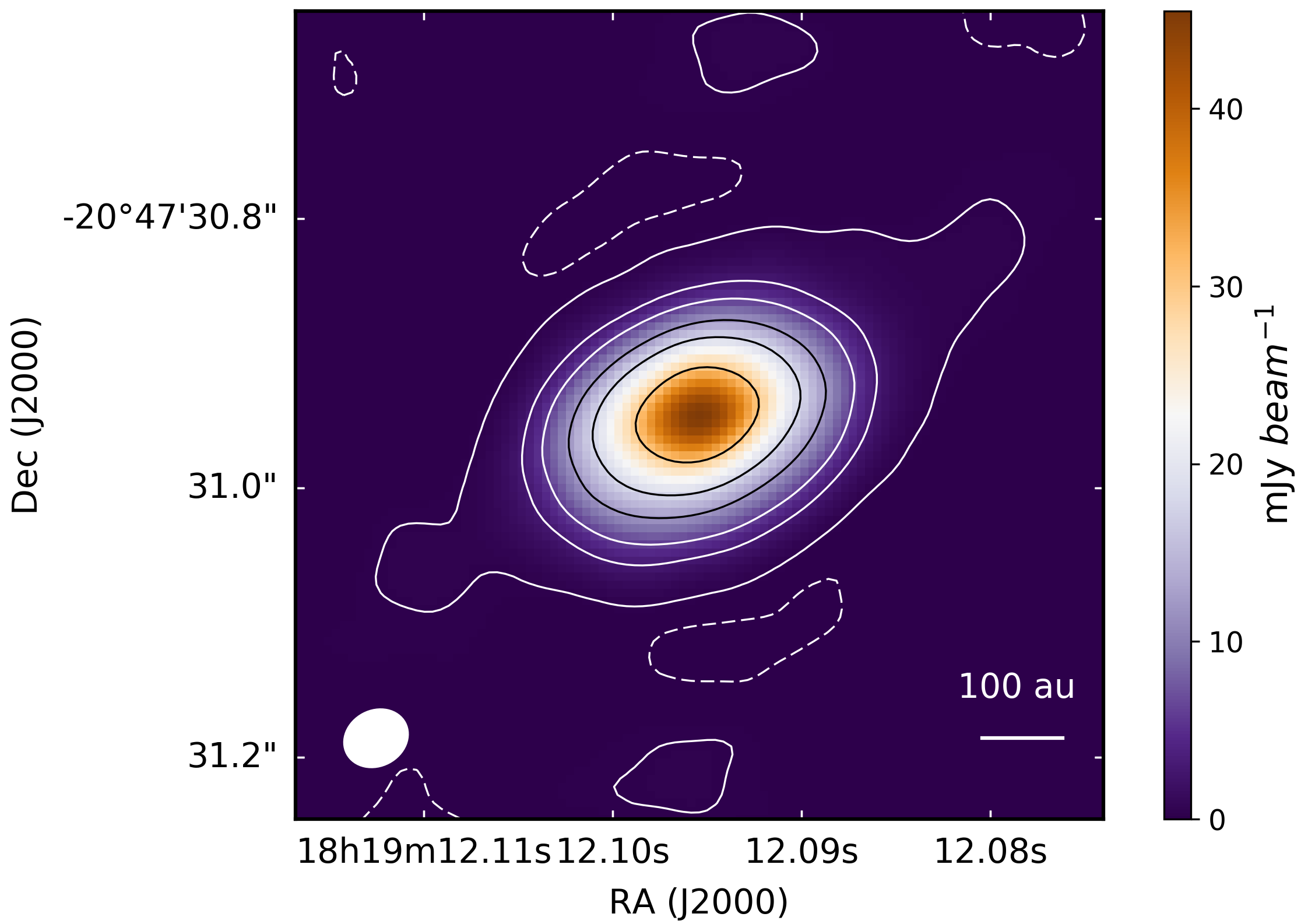}
    \caption{ALMA image at 1.14~mm of the GGD~27-MM1 disk. 
    The contour levels are -5, 5, 50, 100, 200, 300, 500 times the rms noise (0.06 \mJybeam).  The conversion factor from flux density to brightness temperature is $\sim$8.9~K~mJy$^{-1}$~beam.} 
    \label{fig:Real-uv}
\end{figure}

In this section we present the main observational properties that can be used to constrain the parameters of our disk model.
Figure~\ref{fig:Real-uv} shows the ALMA continuum image of the GGD~27--MM1 disk at 1.14~mm. We resolve the disk at this wavelength, obtaining a flux density of 351.30 $\pm$ 0.33~mJy and a peak intensity of 46\mJybeam~ \citep[see also][]{Busquet2019}. The morphology observed at this wavelength is consistent with an inclined disk (49$^{\circ}$; 0$^{\circ}$ for a face-on disk) with a radius of $\sim240$~au and PA$=113^{\circ}$. The brightness temperatures of the disk in the central region reaches a value of $\sim$ 470~K \citep{Girart2018april}. This is an indication of an important source of heating.

Although there are not enough data points at different frequencies with high angular resolution to build the SED of the disk, we have extensive knowledge of the region that provides us with a series of observational constraints regarding the physical parameters of our model, which are discussed below.

%%%%%%%%%%%%%%%%%%%%%%%FIGURES%%%%%%%%%%%%%%%%%%%%%%%%%%%%%%%%%%%%%%%%%

%%%%%%%%%%%%%%%%%%%%%%%%%%%%%%%%%%%%%%%%%%%%%%%%%%%%%%%%%%%%%%%%%%%%%%%%%
%\subsection{Luminosity}
%%%%%%%%%%%%%%%%%%%%%%%%%%%%%%%%%%%%%%%%%%%%%%%%%%%%%%%%%%%%%%%%%%%%%%%%%

\textit{Bolometric luminosity}. The observed value is $\sim$2.0 $\times 10^{4}~L_{\odot}$ for a distance of 1.7~kpc \citep{Gomez2003}. 
As shown in the appendix the source distance is between 1.2 and 1.4 kpc, and therefore the luminosity can be re-calculated to be between $\sim$1.0 $\times 10^{4}$ and $\sim$1.4 $\times$ 10$^{4}$ \lsun. 
This luminosity must be considered an upper limit for the massive protostar GGD~27--MM1 since it comes mainly from the IRAS fluxes that probably encompasses other sources \citep{Fernandez-Lopez2011}.

%%%%%%%%%%%%%%%%%%%%%%%%%%%%%%%%%%%%%%%%%%%%%%%%%%%%%%%%%%%%%%%%%%%%%%%%%
%\subsection{Dynamical mass of the star-disk system }
%%%%%%%%%%%%%%%%%%%%%%%%%%%%%%%%%%%%%%%%%%%%%%%%%%%%%%%%%%%%%%%%%%%%%%%%%

\begin{figure}
    \centering
    \includegraphics[width=0.95\columnwidth]{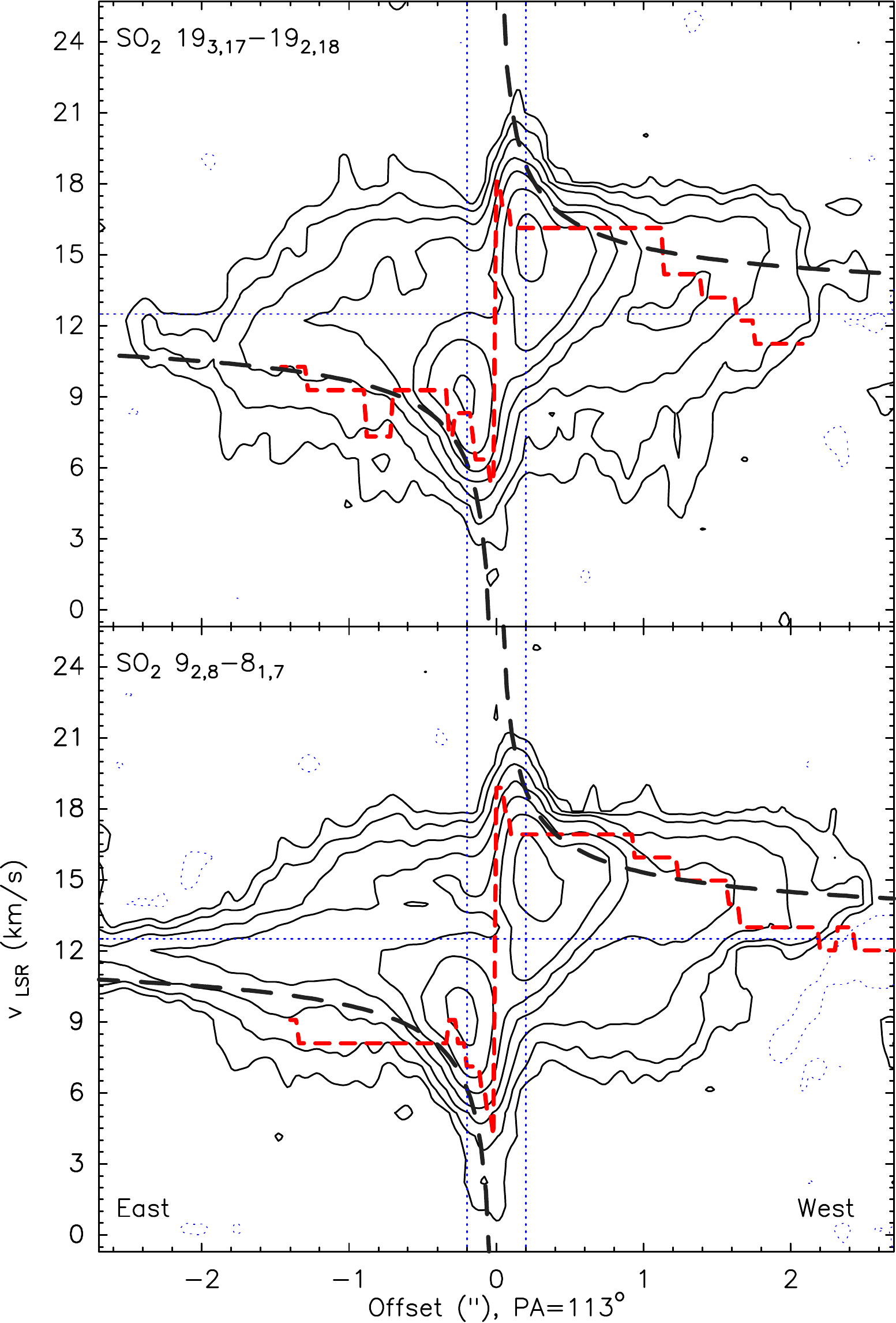}
    \caption{Position-velocity plots along the major axis of the disk for the SO$_2$ 9$_{2,8}$--8$_{1,7}$ and 19$_{3,17}$--19$_{2,18}$ lines. Contours are 2, 4, 8, 16, 32, 64 and 128 times the rms noise of the maps, 1.2 mJy~beam$^{-1}$. The dashed black line shows the expected Keplerian profile for a 25~\msun. The red dashed line shows the best Keplerian fit to the data  using the method proposed by \citet{Seifried16} at 5-$\sigma$. The fit corresponds to a dynamical mass of 30 and 21~\msun\ for SO$_2$ 9$_{2,8}$--8$_{1,7}$ and 19$_{3,17}$--19$_{2,18}$ lines, respectively.
    }
    \label{fig:pvplot}
\end{figure}

\textit{Dynamical mass}. The total dynamical mass of the star-disk system can be obtained from the molecular line emission tracing the gas motions from the disk and assuming that they behave as a rotationally-supported disk (i.e., Keplerian velocity). We used the data cubes for the SO$_2$ 9$_{2,8}$--8$_{1,7}$ and 19$_{3,17}$--19$_{2,18}$ lines (see Section \ref{sec:observation}) to construct position-velocity (PV) maps, centered at the dust peak intensity with a position angle of 113$\arcdeg$, i.e. along the major axis of the disk. Figure~\ref{fig:pvplot} shows the resulting plots. The brightest blueshifted emission appears in the south-east side of the disk, while the redshifted emission arises from the north-west side of the disk. This is in agreement with previous lower angular resolution and less sensitive observations \citep{Fernandez-Lopez2011July, CarrascoGonzalez2012, Girart2017}. We note that there is also significant red/blueshifted emission in the SW/NE side of the disk. This could be an indication of infall motions, although MHD simulation of disk formation and evolution shows that this can be also a projection effect for significant inclinations \citep[e.g.,][]{Seifried16}. 
In order to constrain the dynamical mass from these PV maps, we followed the procedure given by \citet{Seifried16}. This procedure fits the Keplerian profile to the 5-$\sigma$ contour emission, and was tested in synthetic ALMA molecular line PV maps generated from MHD simulations for disks around both low- and high-mass stars. It should best work for lines that fully resolve, spatially and kinematically, the Keplerian profile in the PV maps and for cases not too close to a face-on projection. The best fit obtained yielded an inclination corrected dynamical mass of 31\mm1 and 21\mm1\msun~for the SO$_2$ 9$_{2,8}$--8$_{1,7}$ and 19$_{3,17}$--19$_{2,18}$ lines, respectively. Since the line emission mostly arises from the outskirts of the dusty disk modelled in this paper, we can consider this dynamical mass as the combined mass of the star and the accretion disk.
Thus we explored stellar mass values between 15 and 25\msun~ and keep the total mass (star + disk) close to the dynamical mass. 
%%%%%%%%%%%%%%%%%%%%%%%%%%%%FIGURES%%%%%%%%%%%%%%%%%%%%%%%%%%%%%%%%%%

%%%%%%%%%%%%%%%%%%%%%%%%%%%%%%%%%%%%%%%%%%%%%%%%%%%%%%%%%%%%%%%%%%%%%%%%%
%\subsection{Mass accretion rate}
%%%%%%%%%%%%%%%%%%%%%%%%%%%%%%%%%%%%%%%%%%%%%%%%%%%%%%%%%%%%%%%%%%%%%%%%%

\textit{Mass accretion rate}. \cite{CarrascoGonzalez2012} estimate the mass-loss rate in the jet using the formulation given by \cite{Reynolds1986} (see Eq. \ref{eq:masslossrate}), who model the free-free emission from an ionized jet adopting a power-law dependence with radius \citep[see also][]{Anglada2018}:

\begin{equation}
\begin{array}{c l}
      \left[\frac{\dot{M}_{out}}{M_{\odot}yr^{-1}}\right] = & 1.9 \times 10^{-6}x_0^{-1} \left[\frac{v_{jet}}{1000 km s^{-1}}\right] \left[\frac{S_{\nu}}{mJy}\right]^{0.75} \\
     \times & \left[\frac{\nu}{GHz}\right] ^{-0.45}\left[\frac{D}{kpc} \right]^{1.5}\left[\frac{\theta_0}{rad}\right]^{0.75}
     \label{eq:masslossrate}
\end{array}
\end{equation}

they assume a pure hydrogen jet with constant opening angle $\theta_0$ $\sim$ 19$^{\circ}$ (conical jet), terminal velocity $v_{\mathrm{jet}}$ ($\sim$ 1000 km s$^{-1}$), ionization fraction $x_0$ = 0.1, electron temperature $T_{\mathrm{e}}$=10$^4$~K, and that the jet axis is in the plane of the sky. They obtain a mass-loss rate of $\dot{M}_{\mathrm{out}}$ $\sim$10$^{-5}$ \msunyr~ for a distance of 1.7~kpc, $\simeq$ 8$\times$10$^{-6}$ \msunyr~ for the corrected distance of 1.4~kpc (see appendix). 
This value is in agreement with the value obtained from CO observations of the molecular outflow associated with the jet \citep{Qiu2019}.
Assuming the accretion rate \Macc~of the disk onto the star to be $\sim$10 times larger than the mass-loss rate \citep{Bontemps1996}, the mass accretion rate would be \Macc $\sim$ 8$\times$10$^{-5}$ \msunyr. 
However, \cite{Beltran&deWit2016} (and references therein), obtain a ratio between the mass-loss rate and mass-accretion rate of approximately $\sim$0.3 for disks in young high-mass stars.
In this case the mass accretion rate would be lower, $\sim$ 3$\times$10$^{-5}$ \msunyr.

Even though the mass accretion rate is not well determined using these methods, it allows to define a range of values to explore.
Therefore, in our modeling we explored values of \Macc~in the $\sim$ 1$\times 10^{-5}$ to $\sim$ 2$\times 10^{-4}$\msunyr~range.

%%%%%%%%%%%%%%%%%%%%%%%%%%%%%%%%%%%%%%%%%%%%%%%%%%%%%%%%%%%%%%%%%%%%%%%%%
%\subsection{Stellar parameters}\label{sec:star}
%%%%%%%%%%%%%%%%%%%%%%%%%%%%%%%%%%%%%%%%%%%%%%%%%%%%%%%%%%%%%%%%%%%%%%%%%

\textit{Stellar parameters}. It is expected that most of the dynamical mass (21-31\msun) is stellar, otherwise the disk would be unstable and would show significant asymmetries (e.g., spiral structures) that are not observed with the present data. 
Such massive star, if it were in the main sequence, should develop an \ion{H}{2} region, which is not detected with the present observations (see Sect. ~\ref{sec:ionizedcomponent}).
A possible solution to mitigate this problem is to assume that the star is inflated, with low enough temperature for not producing stellar UV radiation and create an \ion{H}{2} region. 

In that sense, \cite{Hosokawa2009} found a dependence of the protostellar radius with the accretion rate. They obtain that, in general, the higher is the accretion rate, the larger is the stellar radius. Then the protostar has a lower maximum temperature for a certain stellar mass. 
This fact causes a delay in the onset of the main sequence phase and therefore a delay in the formation of an \ion{H}{2} region. 
Thus, because of the high accretion rate estimated, we decided to explore large stellar radii, between 10 and 30 $R_{\odot}$. Because the luminosity is known, this implies temperatures between 12000 and 18000 K. 
This is consistent with \cite{Johnston2013}, who modeled the envelope and disk around the luminous star AFGL 2591-VLA3 (2.3 $\times 10^5$ \lsun), and find a stellar radius of 90 \rsun with a temperature of 13000 K.

%%%%%%%%%%%%%%%%%%%%%%%%%%%%%%%%%%%%%%%%%%%%%%%%%%%%%%%%%%%%%%%%%%%%%
%\subsection{Inner disk radius}
%%%%%%%%%%%%%%%%%%%%%%%%%%%%%%%%%%%%%%%%%%%%%%%%%%%%%%%%%%%%%%%%%%%%%%%%%
\textit{Inner disk radius}. Considering a sublimation temperature of 1400~K for the most refractory grains \citep{D'Alessio2006} and the observed luminosity,
we can locate the sublimation wall from $L_{\mathrm{tot}} = 4 \pi \sigma T^4 R^2$ at $\sim 12$ au of radius.
Because of the high spatial resolution of the observations ($\sim$ 56 au), an inner radius larger than $\sim$ 20 au was discarded. Otherwise we should marginally resolve the inner wall of the disk. 
The sublimation temperature usually is assumed to cover a range from 1200-1500~K, thus we explored an inner radius range between 10 and 20 au instead to just using the 12~au that we calculated (see Table \ref{tab:Parameters}).

%%%%%%%%%%%%%%%%%%%%%%%%%%%%%%%%%%%%%%%%%%%%%%%%%%%%%%%%%%%%%%%%%%%%%%%%%
%\subsection{Settling degree and grain size \label{sec:settling}}
%%%%%%%%%%%%%%%%%%%%%%%%%%%%%%%%%%%%%%%%%%%%%%%%%%%%%%%%%%%%%%%%%%%%%%%%%
\textit{Settling degree and grain size}. We can obtain some constraints about the settling of the larger dust grain at the mid-plane looking at the polarization emission due to scattering from large grains. Based on polarization observations, \cite{Girart2018april} do not find signs of dust settling, and determine a dust maximum grain size ($a_{max}$) from 50 to 500 \mum~(see Section \ref{sec:pola}). We explored in our modeling cases with different settling degree, including the no settling case.

Taking into account the observational restrictions set out above, in the following we proceed to look for the set of parameters that best fit the observations. 
%%%%%%%%%%%%%%%%%%%%%%%%%%%%%%%%%%%%%%%%%%%%%%%%%%%%%%%%%%%%%%%%%%%%%%%%%
\section{The model-fitting procedure \label{sec:procedure}}
%%%%%%%%%%%%%%%%%%%%%%%%%%%%%%%%%%%%%%%%%%%%%%%%%%%%%%%%%%%%%%%%%%%%%%%%%
We computed several grids of models varying the parameters of the disk and the star. Through these grids the parameters were refined until the best-fit model image.

The fitting procedure and analysis was done using Common Astronomy Software Applications (CASA) and Multichannel Image Reconstruction, Image Analysis, and Display \citep[MIRIAD;][]{Sault95} data reduction software packages.

\begin{figure}
\centering
\includegraphics[width=0.99\columnwidth]{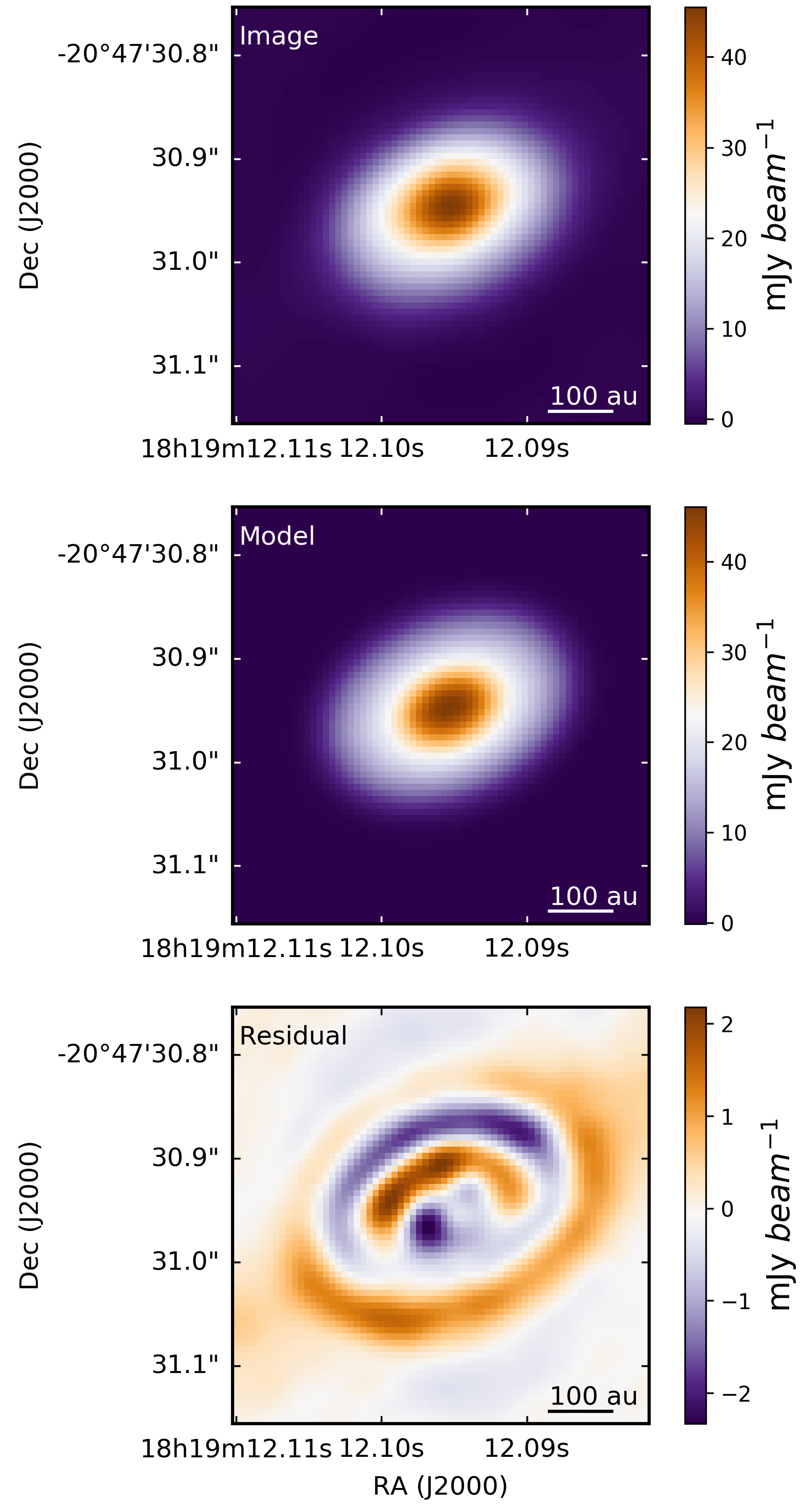}
\caption{Observed ALMA 1.14~mm image (top panel), best fit disk model (middle plane), and the residual (image$-$model) map (bottom panel). 
The conversion factor from flux density to brightness temperature is $\sim$8.9~K~mJy$^{-1}$~beam.
\label{fig:residualpanels}}
\end{figure}

In order to properly compare the model with the ALMA data, synthetic visibilities were computed from the model images. This was done with the CASA {\it simutil} package.
After that, {\it tclean} of CASA was used with the simulated visibilities to create a final image from the model adopting the same cleaning parameters used to create the observed ALMA image. In Fig.~\ref{fig:residualpanels} we show the ALMA image (top panel), the modelled image (middle panel), and the residual map (bottom panel).  
The residual map was obtained by subtracting the model image from the observed one. Therefore positive values in the residuals show regions where the model underestimates the emission.

\begin{figure}
    \centering
    \includegraphics[width=0.99\columnwidth]{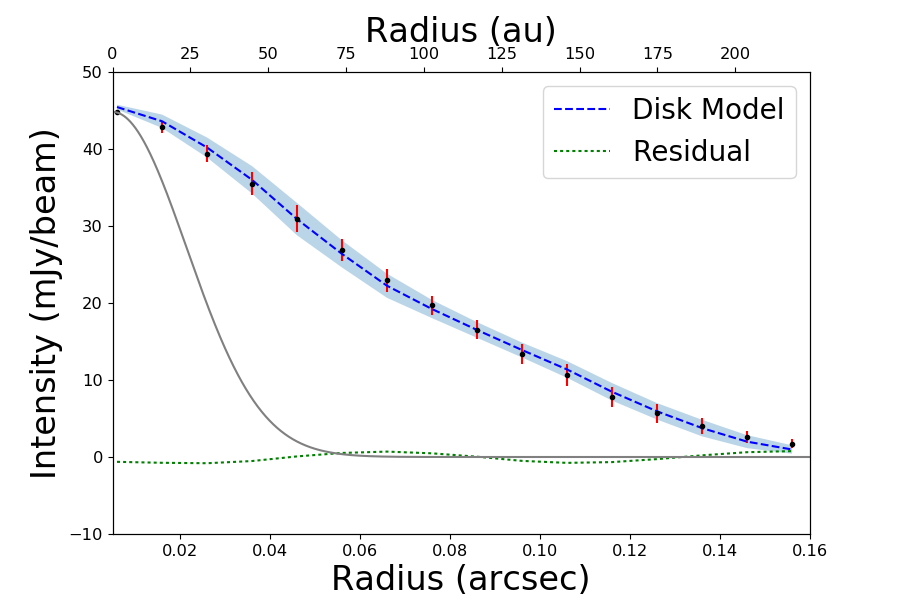}
    \caption{Averaged radial intensity profile at 1.14 mm. Dots represent the averaged value at each radius for the observed image and error bars represent the standard deviation. The blue dashed line is the averaged value at each radius for the disk model and the blue shadow is the standard deviation. The grey line depicts the synthesized beam of the ALMA observations. The dotted green line indicates the residual (difference between observed and modeled profile). \label{fig:Bestfitmodel}}
\end{figure}

The radial intensity profile was computed averaging in concentric ellipses every $0\farcs01$ ($\sim$ 1/4 of the beam) with the inclination adopted in the disk model (see Fig. \ref{fig:Bestfitmodel}). 
The best fit parameters of the disk model were obtained initially by visual inspection of the radial intensity profile and then, from among this first selection we chose the best fit model based on minimum $\chi^2$ of the residual map. 
In the case of the residual map, only pixels inside of ellipse with R=230~au were considered to calculate the $\chi^2$ (see Fig. \ref{fig:Real-uv}) to avoid contamination from the outskirts of the map. Appendix B shows the variation of the disk parameters with $\chi^2$ (leaving the other parameters fixed).
%%%%%%%%%%%%%%%%%%%%%%%%%%%%%%%%%%%%%%%%%%%%%%%%%%%%%%%%%%%%%%%%%%%%%%%%%
\section{Results}\label{sec:results}
%%%%%%%%%%%%%%%%%%%%%%%%%%%%%%%%%%%%%%%%%%%%%%%%%%%%%%%%%%%%%%%%%%%%%%%%%
%%%%%%%%%%%%%%%%%%%%%%%%%%TABLES%%%%%%%%%%%%%%%%%%%%%%%%%%%%%%%

\begin{deluxetable}{l c c }[t]
\tablecaption{Explored parameters range \label{tab:Parameters}}
\tablehead{
\colhead{Parameter} & \colhead{Values} & \colhead{Step} \\
}
\startdata
Star mass ($M_{\odot}$) & 15 -- 25 & 1 \\
Temp. eff. (K) & 12000 -- 18000 & 1000 \\
Star radius ($R_{\odot}$) & 10 -- 30 & 5 \\
Acc. rate (\msunyr) & 1 $\times$ 10$^{-5}$-- 2$\times$ 10$^{-4}$ & 1 $\times$ 10$^{-5}$\\ 
Disk radius (au) & 130 -- 250 & 10 \\
Inner radius (au) & 10 -- 20 & 1 \\
Inclination(degrees) & 44 -- 54 & 1 \\
$a_{max}$ ($\mu$m) & 100, 500 and 1000 & - \\
$\alpha$ & 0.1, 0.01 and 0.001 & - \\
\enddata
\end{deluxetable}

\begin{deluxetable}{l c l}[ht!]
\tablecaption{Best fit model parameters \label{tab:bestmodels}}
\tablehead{
\colhead{Parameter} & \colhead{Value}\\
}
\startdata
$M_*$($M_{\odot}$)    & 20 & fitted\\
$T_{\mathrm{eff}}$ (K)    & 12000 & fitted\\
$R_*$ ($R_{\odot}$)   & 25 & fitted\\
Distance (kpc)      & 1.4 & adopted \\
$\dot{M}$ (\msunyr) & 7 $\times$ 10$^{-5}$ & adopted/refined \tablenotemark{$a$} \\
$M_{\mathrm{disk}}$ ($M_{\odot}$)     & 5 & calculated\\ 
$R_{\mathrm{disk}}$ (au)          & 168 & adopted/refined \tablenotemark{$a$} \\
H$_{100}$ (au)          & 7 & calculated \\
R$_{\mathrm{in}}$ (au)           & 14 & fitted\\
i (degree)         & 49 & adopted/refined \tablenotemark{$a$} \\
$a_{max}$ ($\mu$m) & 500 & fitted \\
$\alpha$ & 0.1 & fitted \\
\enddata
\tablenotetext{$a$}{Parameter with observational constrains and uncertainty.}
\end{deluxetable}

%%%%%%%%%%%%%%%%%%%%%%%%%%%%%FIGURES%%%%%%%%%%%%%%%%%%%%%%%

\begin{figure}
    \centering
    \includegraphics[width=0.99\columnwidth]{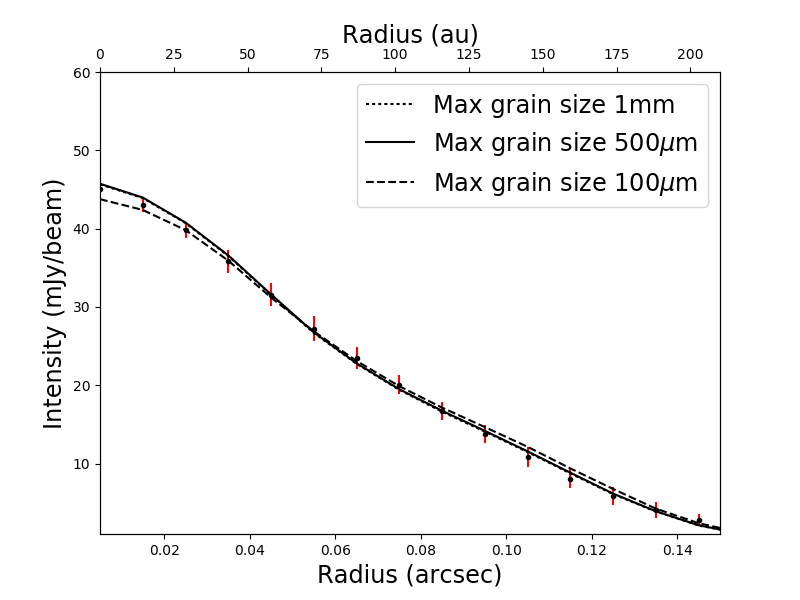}
    \caption{Radial intensity profile for three models with the same parameters given in Table \ref{tab:bestmodels}, except $a_{max}$ of 100 $\mu$m, 500 $\mu$m, and 1 mm.}
    \label{fig:rp_500vs1000}
\end{figure}

\begin{figure*}
    \centering
    \includegraphics[width=\textwidth]{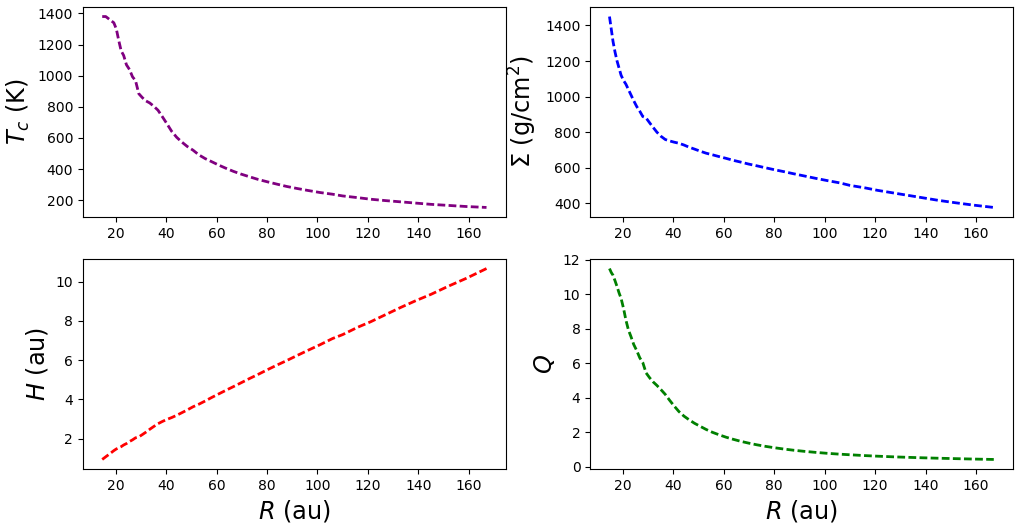}
    \caption{\textit{Upper}: Temperature at the mid-plane (left) and surface density (right). \textit{Bottom}: Scale height, $H$ (left) and Toomre parameter $Q$, see Section \ref{sec:toomre} (right).
    }
    \label{fig:physicalpar}
\end{figure*}

%%%%%%%%%%%%%%%%%%%%%%%%%%%%%%%%%%%%%%%%%%%%%%%%%%%%%%%%%%%%%%%%%%%%%%%%%%%%%%%%%

In this section we present the best fit model obtained after exploring a wide range of values in the space of parameters of the model presented in Section \ref{sec:model}. 
The best fit parameters are shown in Table \ref{tab:bestmodels}. 

We found a massive disk of $\sim$ 5\msun, i.e. 20\% of the stellar mass. The radius of the disk is $\sim$ 170~au with an inclination angle of $\sim$ $49^{\circ}$ (angle between the rotation axis and the line of sight).
The mass accretion rate, which in our model is a very sensitive parameter,
resulted to be $\sim$ 7 $\times 10^{-5}$ \msunyr. The total luminosity is $\sim$ 1$\times$ 10$^4$\lsun which is in agreement with the previous estimation (see Section \ref{sec:HH80--81}). The stellar mass is 20\msun~which together with disk mass (5\msun), is consistent with the estimated dynamical mass for the star-disk system (21--30\msun; see Section \ref{sec:HH80--81}).

Regarding the composition of the disk in terms of grain size, different maximum size of grains according with results found in \cite{Girart2018april} were tested (see Section \ref{sec:HH80--81}). We did not find substantial differences between models with $a_{max}$ of 100, 500~$\mu$m and with 1~mm (see Fig. \ref{fig:rp_500vs1000}). Only for small radii ($<$ 0.05'') the difference between the models is noticeable. 
The grain sizes of the best fit model goes from a minimum value of 0.005~\mum\ to a maximum of 3~\mum\ in the disk upper layer. For those grains settled in the disk mid-plane the grain sizes are between 5~\mum\ and 500~\mum.

The density and temperature profiles for the best fit model are shown in Fig. \ref{fig:physicalpar}. 
We found a flared disk with a maximum scale height of $\sim$13~au.
The disk shows a temperature profile that goes from $\sim$1400~K at the inner edge of the disk to $\sim$150~K at the outer part. 
The small irregularities that can be seen in the internal part of the disk are caused by numerical effects and by the sublimation of the different dust components, which result in a step in the dust opacity, but they do not affect the results.

From the mass surface density ($\Sigma$) and using the total opacity of the model (absorption + self-scattering), $\chi=0.13$ cm$^2/$g (opacity for a dust grain mixture with the physical properties described in Section \ref{sec:model} and with a grain size distribution which assumes grains with a maximum radius of 500 $\mu$m and considers the scattering effects), we obtain an optically thick disk at 1.14~mm for all radii, with $\tau = \Sigma \chi$ ranging from 50 to 170.  In \cite{Busquet2019} they computed the mass of gas and dust of the disk assuming that the 1.14~mm dust continuum emission is optically thin and the temperature distribution is uniform (T$_{\mathrm{d}}$=109~K). They estimated a disk mass for GGD 27 MM1 of $\sim$0.5\msun. This mass could be considered as a lower limit due to the optical thickness of the disk and to the fact that the opacity due to the self-scattering is not considered. 
We fitted the density and the temperature profiles with a power law functions ($\Sigma(R) \propto R^p$, $T(R) \propto R^q$) using the method of the minimum mean squared error (MSE). Here we present the coefficient and the power index.     
Equations \ref{eq:powerlaw} show the behaviour of the surface density, $\Sigma$($R$), 
and the temperature at the mid-plane, $T_{\mathrm{c}}$($R$), approximated as power laws. The MSE of the fits are 0.023, and 0.034 respectively. 

\begin{equation}
\begin{array}{rl}
    \frac{\Sigma(R)}{\left[\mathrm{g/cm^2}\right]} \sim & 500 \left(\frac{R}{\left[\mathrm{100~au}\right]}\right)^{-0.5}  \\
%    T_{zs}(R) = & 4.5 \times (R/R_0)^{-1}  \\
    \frac{T_c(R)}{[\mathrm{K}]} \sim & 300 \left(\frac{R}{\left[\mathrm{100~au}\right]}\right)^{-1}
\end{array}
\label{eq:powerlaw}
\end{equation}

%%%%%%%%%%%%%%%%%%%%%%%%%%%%%%%%%%%%%%%%%%%%%%%%%%%%%%%%%%%%%%%%%%%%%%%%%%%%%%%%%
\section{Discussion}
%%%%%%%%%%%%%%%%%%%%%%%%%%%%%%%%%%%%%%%%%%%%%%%%%%%%%%%%%%%%%%%%%%%%%%%%%%%%%%%%%
In this work we have studied the ALMA image at 1.14~mm of the circumstellar disk around GGD~27 -MM1. 
We found a compact source coming from the inner radius ($\sim$ 4~mas/5.6~au) probably due to ionized gas.
By modeling the dust continuum emission from the disk (with the compact source previously subtracted), we found a massive ($\sim$ 5 \msun) and compact ($\sim$ 170~au) disk. 
In the following we discuss the implications of our results, while also analyzing the gravitational stability of the disk.

%%%%%%%%%%%%%%%%%%%%%%%%%%%%%%%%%%%%%%%%%%%%%%%%%%%%%%%%%%%%%%%%%%%%%%%%%%%%%%%%%
\subsection{Ionized component \label{sec:ionizedcomponent}}
%%%%%%%%%%%%%%%%%%%%%%%%%%%%%%%%%%%%%%%%%%%%%%%%%%%%%%%%%%%%%%%%%%%%%%%%%
The compact source reported in Sect.~\ref{sec:compact} appears unresolved in the different images obtained with a minimum visibility radius of 4000~k$\lambda$ (Fig. \ref{fig:ALMcompact}). The Gaussian fits in the visibility domain indicate that the source has a radius of $\sim$5.6~au and a brightness temperature of $\sim$10$^{4}$~K.
At such high temperature, the dust grains should be sublimated. Indeed, the expected temperature for the sublimation of silicates is $\sim$1400~K \citep{D'Alessio2006}. Therefore, the most plausible explanation is that this compact emission is tracing ionized gas, either from an incipient and extremely compact \ion{H}{2} region or from the base of the
HH~80--81 thermal radio-jet, best traced at cm-wavelengths \citep[e.g.,][]{CarrascoGonzalez2012}. 
In fact, from the peak flux density measured at 1.3~cm at the center of the radio-jet \citep[$\sim$1~mJy;][]{CarrascoGonzalez2012}, and the flux density measured at 1.14~mm $\sim$19~mJy (see Sec. \ref{sec:compact}), we estimated a spectral index of $\alpha$ $\sim$1.2 between these two frequencies. This value is within the range of spectral indices measured in thermal radio-jets associated with YSOs \citep[e.g.,][]{Anglada2018}. 
Higher resolution and multi-frequency observations would help determine the nature of this compact component in a conclusive way. 

%%%%%%%%%%%%%%%%%%%%%%%%%%%%%%%%%%%%%%%%%%%%%%%%%%%%%%%%%%%%%%%%%%%%%
\subsection{Restrictions from polarization \label{sec:pola}}
%%%%%%%%%%%%%%%%%%%%%%%%%%%%%%%%%%%%%%%%%%%%%%%%%%%%%%%%%%%%%%%%%%%%%%%%%
An important source of information about the composition of the disks in terms of grain size, grain shape, and grain distribution comes from polarimetric observation. 

Polarization data allows to constrain the dust distribution in the disk (settling) and the maximum dust grain size. 
\cite{Girart2018april} based on polarization models of \cite{Yang2017} conclude that dust settling has not yet occurred . \cite{Yang2017} compare two models with different thickness of the layer of large grains which are responsible for the scattering, and propose two models in which large grains can be found up to 0.1~$H'$ and 1~$H'$, where $H'$ is the (dust) hydrostatic scale height, $H'(R)=H'_0 (R/R_c)^{1.5-q/2}$ being $R_{\mathrm{c}}$ a characteristic radius of the disk (dust) density distribution, and \textit{q} the temperature power law index.

Unlike the models of these authors, our models consider the dust settling changing the dust-to-gas mass ratio between two populations through the $\epsilon$ parameter as explained in Section \ref{sec:model}. Small grains are in the upper layer and big grains are in the mid-plane. The position of the border between these two populations can be set through the parameter $Z_{\mathrm{big}}$. 

In this work our best fit model has $\epsilon$=1 (no dust settling) but it still has two dust populations with the border located at $Z_{\mathrm{big}}$=0.1$H$, where $H$ is the pressure scale height of the gas at the mid-plane temperature (see Section \ref{sec:model}).

We also reproduced the scenario suggested by \cite{Yang2017}
to reproduce the polarization pattern observed by \cite{Girart2018april}. To do so, we tested a model with Z$_{\mathrm{big}}$ =1 $H$ and $\epsilon$ =1.
In this case we obtained a more massive disk (7 \msun) with larger mass accretion rate (1 $\times$ 10$^{-4}$ \msunyr) than the model with $Z_{\mathrm{big}}$=0.1$H$ ($M_{\mathrm{disk}}$ $\sim$ 5 \msun~and \Macc $\sim$ 7 $\times$ 10$^{-5}$ \msunyr; see Table \ref{tab:bestmodels}), our fiducial best fit model, which shows a lower value of $\chi^2$. 

Furthermore, we have compared two- and one-population models. We tested models with a single dust grain population with a maximum grain size of 500~\mum.
Due to the lower opacity to the stellar radiation of this dust population, the one-population models would require significantly higher disk masses, even comparable to the stellar mass. These masses are inconsistent with our estimated dynamical mass (see Section \ref{sec:HH80--81}, indicating that at least two dust populations are needed to fit the observations.

In summary, we have considered the results presented in \cite{Girart2018april} concerning the size and distribution of grains. Moreover, taking into account the conceptual differences presented above ($H$, $H'$), we have explored values adjacent to them. We did not find substantial differences between both scenarios and therefore we can not favor either.  

%%%%%%%%%%%%%%%%%%%%%%%%%%%%%%%%%%%%%%%%%%%%%%%%%%%%%%%%%%%%%%%%%%%%%%%%%%%%%%%%%
\subsection{Stability of the disk \label{sec:toomre}}
%%%%%%%%%%%%%%%%%%%%%%%%%%%%%%%%%%%%%%%%%%%%%%%%%%%%%%%%%%%%%%%%%%%%%%%%%%%%%%%%%
We have quantified the Toomre parameter of our model to check the stability of the disk against self-gravity perturbations. Figure \ref{fig:physicalpar} (bottom right) shows the Toomre parameter $Q$ (Eq. \ref{eq:tommre}) for a Keplerian disk evaluated at the disk mid-plane temperature.

\begin{equation}
    \begin{array}{cl}
       Q =  & \frac{ c_s \Omega}{\pi G \Sigma}  \\
        
       \Omega =  & \left( \frac{G M_{total}}{R^3} \right)^{1/2}, \\ 
    \end{array}
    \label{eq:tommre}
\end{equation}
where $c_s$ is the sound speed at the mid-plane temperature, $\Omega$ the Keplerian angular velocity, $\Sigma$ the surface density of the disk, and $G$ the gravitational constant. 

We adopted the surface density and the temperature at the mid-plane as a function of the radius of our best fit model.
The stability condition ($Q>1$) is satisfied up to a radius $R_{\mathrm{disk}}\sim100$~au. A similar result has been found by \cite{Maud2019} in the massive O-type protostar G17.64+0.16, who reported a massive and stable disk for $R_{\mathrm{disk}}$ $\leq$ 150~au. Furthermore, \cite{Meyer2018} and \cite{Takahashi2016} propose an update of the Toomre criterion in which the only necessary condition for disk instability is Q$<$0.6.

In addition, our best fit model satisfies the conditions proposed by \cite{Forgan_2016} in order to apply the $\alpha$-viscosity prescription model in self-gravitating disks; that is, $H/R$ $<$ 0.1 (see Fig. \ref{fig:physicalpar}) and $M_{\mathrm{disk}}$/$M_*$ $<$ 0.5. 

We note that due to the relatively high mass of the disk of our best model (5\msun), self-gravity might play a non-negligible role.
In order to explore the potential effects of including self-gravity, we compared our best fit model to the hydrodynamic simulations (including self-gravity) of massive star formation performed by \cite{Kuiper2018}.
Our model shows very similar results in terms of disk mid-plane temperature and disk's aspect ratio to this model \citep[Fig. 6;][]{Kuiper2018}, which indicates that including self-gravity would not change our results significantly.
%%%%%%%%%%%%%%%%%%%%%%%%%%%%%%%%%%%%%%%%%%%%%%%%%%%%%%%%%%%%%%%%%%%%%%%%%%%%%%%%%
\subsection{Residual map }
\begin{figure}
    \centering
    \includegraphics[width=0.98\columnwidth]{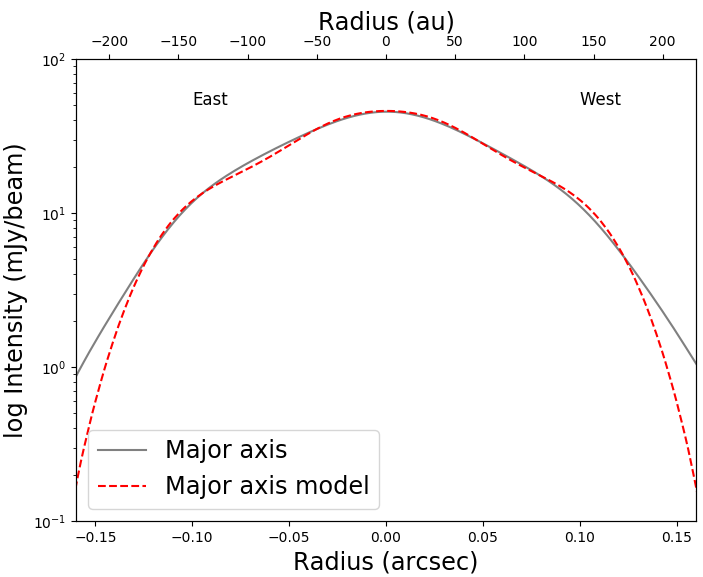}
    \includegraphics[width=0.98\columnwidth]{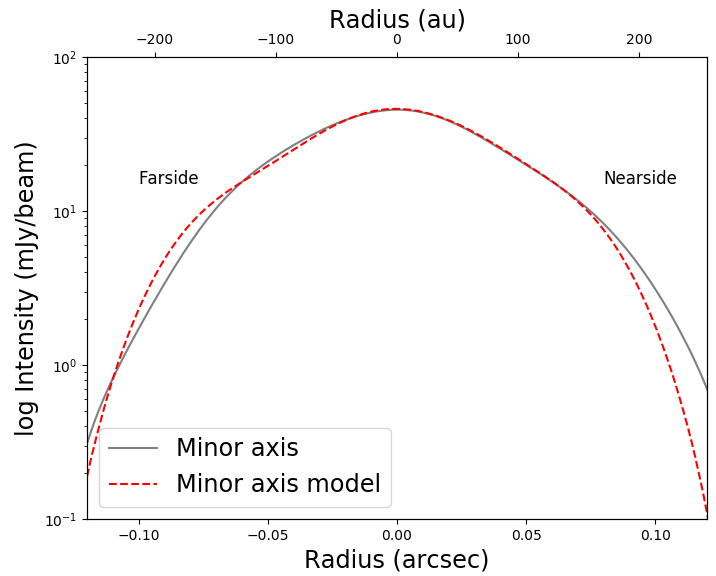}
    \caption{Cut along the major (top panel) and minor (bottom panel) axis. Solid grey and dashed red lines represent the observed image and model, respectively. The physical space scale (au) is corrected by inclination.}
    \label{fig:majorminorcuts}
\end{figure}

\begin{figure*}
    \centering
    \includegraphics[width=0.9\textwidth]{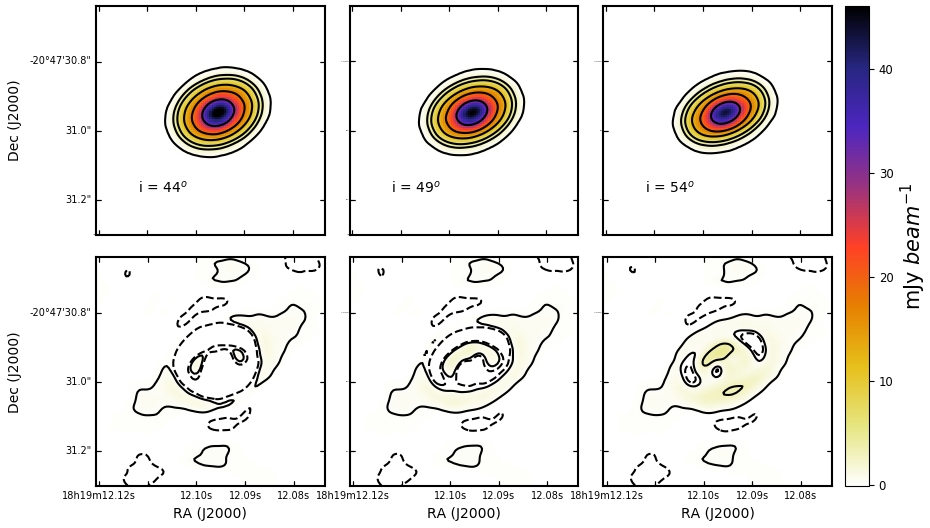}
    \caption{Models (top panels) and residuals (bottom panels) for the best fit model 
    varying the inclination. Left panels correspond to $i$=44$^{\circ}$, middle panels to $i$=49$^{\circ}$, and right panels to $i$=54$^{\circ}$. The contours levels are -5, 5, 50, 100, 200, 300, and 500 times the rms of the observed image (0.06 \mJybeam). }
    \label{fig:panel_grid_i}
\end{figure*}

%%%%%%%%%%%%%%%%%%%%%%%%%%%%%%%%%%%%%%%%%%%%%%%%%%%%%%%%%%%%%%%%%%%%%%%%%%%%%%%%%
The values of the residual map are low when compared with the observed (residuals are below 5$\%$ of the peak intensity), but significant with respect to the rms noise level ($\sigma$ $\sim$0.06 mJy beam$^{-1}$), with an intensity range between $\sim$ 2 and $-$2 \mJybeam~(see Fig. \ref{fig:residualpanels}). 
The main differences between the observed and modeled images 
arise from the outer parts of the disk, beyond 150~au, with an excess and a deficit of emission. This is illustrated in Fig. \ref{fig:majorminorcuts}, where we show flux density-position cuts along the major and minor axis of the disk (observed and modeled).
We also verified that the asymmetries were not caused by a mismatch in the inclination of the system or a shift in the disk centers. In Fig. \ref{fig:panel_grid_i} we present the model image and the residual map of the best fit model ($i$=49$^{\circ}$; central panels) together with two models in which the inclination around the best fit has been modified ($i$=44$^{\circ}$ and 54$^{\circ}$). 
In addition, we show in
Fig. \ref{fig:xi2} 
the $\chi^2$ as a function of the inclination for the best fit model varying the inclination angle in a range of 10$^{\circ}$ centered at 49$^{\circ}$. 

Although we cannot discard intrinsic asymmetries within the disk, a plausible explanation for this excess/defect of emission at the outer parts of the disk could be a
small mismatch with the flaring angle and/or settling of the disk. 
To discriminate between possible intrinsic asymmetries and modeling, we would need ALMA high-angular observations at other frequencies as it has been carried out in low-mass YSOs \citep[e.g.,][]{CarrascoGonzalez2019,Macias2019}. 

%%%%%%%%%%%%%%%%%%%%%%%%%%%%%%%%%%%%%%%%%%%%%%%%%%%%%%%%%%%%%%%%%%%%%%%%%%
\subsection{Mass accretion rate and evolutionary stage}
%%%%%%%%%%%%%%%%%%%%%%%%%%%%%%%%%%%%%%%%%%%%%%%%%%%%%%%%%%%%%%%%%%%%%%%%%%
\cite{Trinidad2003} used the D'Alessio model to fit the SED of AFGL 2591 VLA3, a massive disk-star system located in the Cygnus X region. 
They find that for all models, the main heating source was the stellar irradiation for radii larger than 20~au and the viscous dissipation for smaller radii. Furthermore, $\Sigma$ $\sim$\Macc/$\alpha$ for radius larger than 20~au. 
In this scenario, they find a family of models that could explain the observed SED. This family of models correspond to a constant value of \Macc/$\alpha$, showing that the SED does not change as long as this ratio is maintained. 

\begin{figure}
    \centering
    \includegraphics[width=0.99\columnwidth]{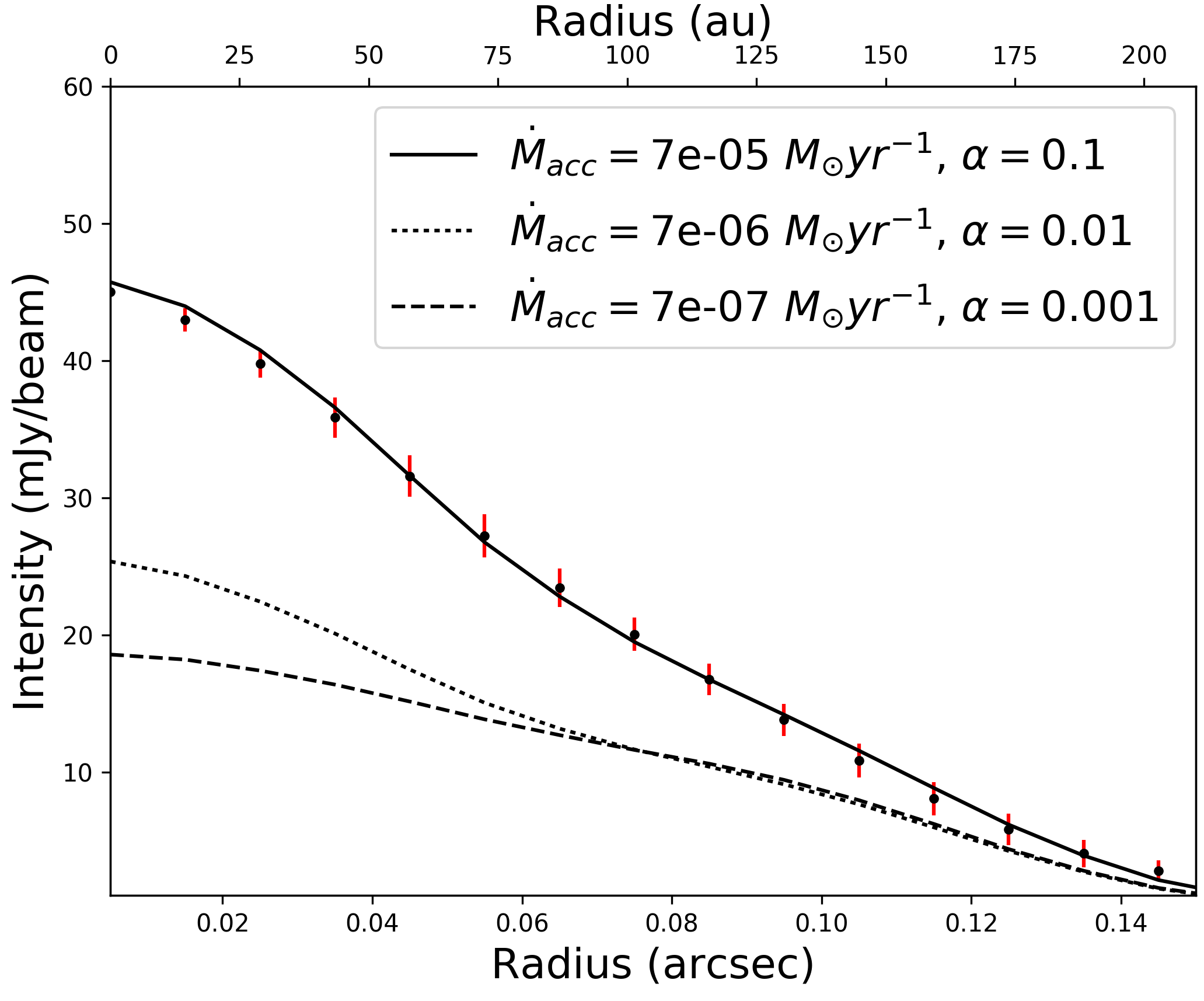}
    \caption{Averaged radial intensity profile of the best fit model ($R_*$=25 $R_{\odot}$, $M_*$=20\msun, $R_{\mathrm{disk}}$=170~au, $i$=49$^{\circ}$, $T_{\mathrm{eff}}$=12000~K, distance=1.4~$kpc$, $R_{\mathrm{in}}$=14~au, $a_{max}$ 500 $\mu$m) varying \Macc and $\alpha$. Solid line: Disk mass accretion rate 7$\times$10$^{-5}$\msunyr, $\alpha$=0.1, $M_{\mathrm{disk}}$=5\msun. Dotted line: Disk mass accretion rate 7$\times$10$^{-6}$\msunyr, $\alpha$=0.01, $M_{\mathrm{disk}}$=8\msun. Dashed line: Disk mass accretion rate 7$\times$10$^{-7}$\msunyr, $\alpha$=0.001, $M_{\mathrm{disk}}$=9\msun. }
    \label{fig:maccalpha}
\end{figure}

Our best fit model yields a rate \Macc/$\alpha$ $\sim$1$\times$10$^{-4}$ \msunyr. We tested if the radial intensity profile would be significantly affected by varying the accretion rate and alpha, while keeping \Macc/$\alpha$ constant.
As we show in Fig. \ref{fig:maccalpha}, the radial intensity profile at 1.14~mm is dramatically affected when the disk mass accretion rate is changed although the ratio \Macc/$\alpha$ stays constant.
The main reason for this difference is because of the effects of the irradiation from the accretion luminosity. 
Therefore by not having this degeneration we are able to constrain the \Macc.

We would like to point out that the accretion rate could be variable with time. In fact, \cite{Marti1998}, based on multi-epoch VLA continuum observations, report a flux density decay of the two inner condensations in the HH~80--81 thermal radio jet. Such a flux density decay could be attributed to changes in the mass accretion rate, being higher in the past.
Furthermore, GGD~27--MM1 is a young source that has a faint envelope and probably an incipient (hyper-compact) \ion{H}{2} region. Some studies estimate that the timescale for the development of \ion{H}{2} regions, with an accretion rate in the range $\sim$ 10$^{-4}$--10$^{-3}$ \msunyr, is 
$\sim$10$^5$ years \citep[e.g.,][]{Osorio1999, Cesaroni2005}. Considering the constant accretion rate of our best fit model (~7$\times$10$^{-5}$\msunyr), for it to reach a star mass of 25\msun, its age would be $\sim$4$\times$10$^5$~yr, which is larger than the development time of an \ion{H}{2} region. Thus, higher accretion episodes in the past are necessary to explain the present situation.

The GGD~27 complex includes two compact cores, MM1 and MM2, separated by $\sim$7'' ($\sim$ $10000$~au).  
\citet{Fernandez-Lopez2011} estimated masses of MM1 and MM2 at different scales (see their Table~6). These authors show that while in MM2 most of the mass ($\sim$75$\%$) is found at envelope scale, in MM1 $\sim$70$\%$ of the mass is already at disk scale. This fact would place MM1 in a more evolved stadium  than MM2, being MM1 equivalent to a Class I low-mass star.

\subsection{Temperature distribution in the GGD~27-MM1 disk:
comparison with disks around low- and intermediate- mass stars and
implications on the water snow line \label{subsec:snow-line}}

In flared $\alpha$-irradiated  disk models, the temperature varies as a function of both the radial distance to the star and the height above the disk mid-plane. The mid-plane temperature decreases with increasing radius, typically as $T_c \propto R^{[-0.5,-1]}$ \citep[e.g.,][]{D'Alessio1998}. In the GGD~27-MM1 disk we found $T_c \propto R^{-1}$ (see Section \ref{sec:results}).

Furthermore, because of the flared morphology of the disk, its surface is heated directly by the radiation from the star and the accretion shock, while the inner layers are heated by viscous dissipation, which heats mainly the disk mid-plane. 
The energy released by the star and the accretion shock generally exceeds the one released by viscous dissipation. As a result of these heating mechanisms, the surface of the disk is warmer than the regions closer to the mid-plane, but the temperature gradient becomes smoother, or even reversed (temperature increasing with decreasing height), near the mid-plane.
In the latter case, the minimum temperature is not reached at the mid-plane but above it.

For disks in low-mass stars this vertical inversion only happens typically at radii $\leq$ 1 au and heights $\leq$ 1 au \citep{D'Alessio1997, D'Alessio1998}. According to our modeling, in the case of the GGD~27--MM1 disk the vertical inversion of temperature
occurs at all radii; for instance, near the star, at $R_{\mathrm{disk}}\simeq$ 30 au, the inversion occurs at a height of $\sim$ 5 au and at a temperature of $\sim$ 240 K, while in the outermost regions, $R_{\mathrm{disk}}\simeq$ 150 au, the inversion occurs at a height of $\sim$ 30 au and at a temperature of $\sim$120 K.

Typical disks around low-mass stars reach very low temperatures \citep[$\sim$ 20--30~K at $\sim$ 150~au;][]{D'Alessio1997, Menshchikov1999, Fogel2011, Tobin2018}. Disks around intermediate-mass stars have slightly higher values of the minimum temperature \citep[$\sim$ 30--40~K at $\sim$ 150~au;][]{Osorio2014}. In contrast, according to our modeling, the GGD~27--MM1 disk is significantly warmer. Even at large distances from the star, close to the edge of the disk ($R_{\mathrm{disk}}\simeq$ 170 au), its mid-plane temperature remains above $\sim$140 K (see Fig. 9) and at this radius the minimum temperature is $\sim$ 115 K at a height of $\sim$ 30 au. High temperatures have also been reported for disk candidates around other massive protostars (mid-plane $T_c \simeq$ 200~K at $R_{\mathrm{disk}}\simeq$ 200~au;
\citealt{Chen2016}), using radiative transfer models which take into account radial and vertical temperature gradients.
However in most cases, the temperatures have been inferred from
vertically isothermal models that do not provide the temperature of the surface layer; also, it is unclear if some of the disk candidates are indeed real accretion disks or just elongated structures, since they are extremely large.

Due to the elevated temperatures of GGD~27--MM1 disk, condensation fronts,
known as snow lines, of major volatile such as carbon monoxide,
carbon dioxide and methane, are not expected to be present since these
species sublimate at temperatures considerably lower ($\sim$ 20--40 K,
\citealt{Zhang2015}) than the temperatures we find in the GGD~27--MM1 disk.
Nevertheless, water ice sublimates at temperatures above 100 K,
hence it is possible that the water snow line is present in the
outer regions of GGD~27--MM1 disk. Since the water ice sublimation
temperature depends on the density \citep[e.g.,][]{Sandford&Allamandola1993, Osorio2009}, taking into account the density and temperature
distributions obtained in our modeling  of the GGD~27--MM1 disk (see Section \ref{sec:results}), we estimate that the water snow line would be located near the edge of the disk, at a radius of $\sim$ 170 au. At this radius, densities are
$1\times10^9$--$2\times10^{11}$ cm$^{-3}$ within a height of 30~au of the mid-plane, implying a range of water sublimation temperatures of 120--130 K, which is similar to the range of temperatures of $\sim$ 120--150 K predicted by our model. 

It has been thought that water snow lines are important because they can trigger the growth of grains to pebbles and lastly to planets.
They are also important because they mark the border between rocky planets formed inward of this line and giant gas planets formed outside.
In the case of the GGD~27--MM1 disk, we expect the formation of gas planets to be hindered by the high temperatures of the disk, being restricted to radii near the disk edge. Consequently, we speculate that if the formation of gas planets were to occur in disks around massive protostars, in general, they would be formed at distances around hundreds of au. It is unclear, however, whether such giant planets could survive after the onset of an \ion{H}{2} region around the massive star.

One should keep in mind that the planetary formation process in high-mass protostars, if it takes place, must be fast because the time scales for the formation of these stars are shorter than for low-mass stars. In principle, the time scale for planet formation is expected to be of the order of that of mass exchange in the accretion disk, roughly estimated as the disk mass divided by the accretion rate, resulting in $7\times10^4$ yr for the disk of GGD~27--MM1. This time scale is shorter than the values typically estimated for the grain coagulation process ($\sim 10^6$ yr, \citealt{Testi2014}), required for particles to become cores and eventually planetesimals ending in planets. However, some additional factors should be also taken into account: (i) the full disk lifetime is larger than the time scale of mass exchange in the disk, resulting in more available time for the final planetary mass to be assembled. In particular, in the GGD~27--MM1 disk, where infall is still significant, the disk is still being replenished with new material from the surrounding envelope that makes its life longer; (ii) at later stages, hydrodynamic models show that the accretion of material onto planet embryos can largely exceed the accretion onto the star itself \citep{zhu2011}, speeding up the planet formation; (iii) the density of particles in massive disks is much higher than in low-mass disks. Thus, the density of planetesimals would be higher and one would expect planet assembling to be faster than in the low-mass case, analogously to what happens in the formation of the star itself, which is a faster process for high-mass stars; (iv) lastly, for low-mass stars there is both theoretical \citep{Lambrechts2012} and observational evidence (HL Tau: \citealt{CarrascoGonzalez2016}; TMC 1A: \citealt{Harsono2018}) suggesting that planet formation starts very early in the star-formation process. These results imply that the planetary formation process might be faster than initially thought and compatible with the time scales of massive star-formation.

Obtaining observational constraints on the location of the
water snow line in the GGD~27--MM1 disk would be of major importance to inform about the plausibility of gas planet formation around this massive protostar. 
We note that in low-mass protostars, water snow lines are difficult to detect because they are commonly located at very small distances, of only a few au, from the star, requiring a very high angular resolution to detect them. 
\cite{Cieza2016} observed V883 Ori during an outburst, when an increase in luminosity drove the water snow line out to more than 40 au ($\sim 0.1''$ at the distance of Orion), making the detection feasible. Since in the GGD~27--MM1 disk the water snow line is expected to be located at a radius
of $\sim$ 170 au, resulting in a similar angular separation, $\sim0.1''$, at the distance of GGD~27--MM1, it would not require of a stellar outburst, as in the case of low-mass protostars, to become detectable. 
High angular resolution ALMA observations of the GGD~27--MM1 disk at several frequencies
could help to constrain the presence of the water snow line by looking for spatial variations of the dust optical depth \citep[e.g.,][]{Cieza2016}.
%%%%%%%%%%%%%%%%%%%%%%%%%%%%%%%%%%%%%%%%%%%%%%%%%%%%%%%%%%%%%%%%%%%%%%%%%%%%%%%%%
\section{Conclusions}
%%%%%%%%%%%%%%%%%%%%%%%%%%%%%%%%%%%%%%%%%%%%%%%%%%%%%%%%%%%%%%%%%%%%%%%%%%%%%%%%%
We used ALMA continuum observations at 1.14~mm, obtained with an angular resolution of $\sim$40~mas, to model the accretion disk around the central massive early B-type protostar exciting the powerful HH~80--81 radio jet using $\alpha$-viscosity prescription. We found an enclosed mass of 21--30\msun, of which 5--7\msun~can be attributed to the disk. This mass is consistent with the derived dynamical mass of $31\pm1$\msun~and $21\pm1$\msun~for the SO$_2$\,9$_{2,8}$--8$_{1,7}$ and 19$_{3,17}$--19$_{2,18}$ lines, respectively. The radius of the disk is $\sim$170~au, with an inclination angle of 49\degr.
We compared the physical structure, temperature and density profiles, obtained with our model with power law functions, showing that the GGD~27--MM1 system is a potential template for future similar studies in other high-mass protostars. In particular, we obtained a flared disk with a maximum scale height of $\sim$13~au, and a temperature profile that ranges from $\sim$150~K at the outskirts of the disk up to $\sim$1400~K at the inner edge of the disk. The analysis of the Toomre $Q$ parameter, evaluated at the disk mid-plane temperature, indicates that the disk is stable up to a radius $R_{\mathrm{disk}}\simeq$100~au.
This work shows that the D'Alessio models can be used as a first approximation in the modeling of accretion disks around massive protostars, providing in addition several observational predictions.

We also reported the presence of an unresolved compact source at the center of the accretion disk, with a radius of 4~mas ($\sim$5.6~au at the source distance of 1.4~kpc) and a brightness temperature of $\sim$10$^4$~K, most likely tracing ionized gas. 
The origin of this compact source is uncertain,  it could arise from an incipient, extremely compact \ion{H}{2} region or from the base of the HH~80--81 radio jet. Observations at higher angular resolution would help to determine the nature of this compact source.

Finally, we have estimated a distance of 1.2-1.4~kpc to the GGD\,27 star-forming region based on the Gaia DR2 catalogue combined with near-IR polarimetric data of the YSOs in the region and the extinction maps. 

%%%%%%%%%%%%%%%%%%%%%%%%%%%%%%%%%%%%%%%%%%%%%%%%%%%%%%%%%%%%%%%%%%
\begin{acknowledgments}
%%%%%%%%%%%%%%%%%%%%%%%%%%%%%%%%%%%%%%%%%%%%%%%%%%%%%%%%%%%%%%%%%%

We are grateful to Nuria Calvet for making D'Alessio's code available to us and for very useful discussions. We would also like to deeply thank Guillem Anglada for his critical comments on this work that have undoubtedly improved it very substantially. We thank our anonymous referee for her/his very helpful comments and suggestions to improve our manuscript.
We also thank Rosine Lallement for the support on {\it STILISM}.
This paper makes use of the following ALMA data: ADS/JAO.ALMA\#2015.1.00480.S. ALMA is a partnership of ESO (representing its member states), NSF (USA) and NINS (Japan), together with NRC (Canada) and NSC and ASIAA (Taiwan) and KASI (Republic of Korea), in cooperation with the Republic of Chile. The Joint ALMA Observatory is operated by ESO, AUI/NRAO and NAOJ.
This work has made use of data from the European Space Agency (ESA) mission
{\it Gaia} (\url{https://www.cosmos.esa.int/gaia}), processed by the {\it Gaia}
Data Processing and Analysis Consortium (DPAC,
\url{https://www.cosmos.esa.int/web/gaia/dpac/consortium}). Funding for the DPAC
has been provided by national institutions, in particular the institutions
participating in the {\it Gaia} Multilateral Agreement.
N. A.-L, M.O., J.M.G,% G.A., 
 G.B., R.E., and J.M.T.  are supported by the Spanish grant 
AYA2017-84390-C2-R (AEI/FEDER, UE). R.E. is also supported by MDM-2014-0369 of ICCUB (Unidad de Excelencia ``Mar\'ia de Maeztu"). 
M.O. acknowledges financial support from the State Agency for Research of the Spanish MCIU through the ``Center of Excellence Severo Ochoa'' award for the Instituto de Astrof\'isica de Andaluc\'ia (SEV-2017-0709).
S.C. acknowledges support from DGAPA, UNAM and CONACyT, M\'exico. M.F-L. acknowledges support from the LACEGAL project which has received funding from the European Union's Horizon 2020 Research an Innovation Program under the Marie Sklodowska-Curie grant agreement No 734374. R.G.M. acknowledges support from UNAM PAPIIT project IN 104319. J.K. is supported by MEXT KAKENHI grant number 19K14755.   
\end{acknowledgments}

%%%%%%%%%%%%%%%%%%%%%%%%%%%%%%%%%%%%%%%%%%%%%%%%%%%%%%%%%%%%%%%%%%%%%%%%%%%%%%%

\software{CASA \citep{McMullin2007}, MIRIAD \citep{Sault95}, PYTHON, STILISM}

\appendix \label{sec:appendix}

\section{The distance to the LDN 291 cloud}\label{Appendix}
The GGD~27 nebulosity and the objects associated to the region  \citep[e.g., HH~80-81, and the HH~80N star forming core][]{Heathcote98, Girart1994, Girart2001, Masque2011, Masque2013} are located in the LDN 291 large molecular cloud complex (including the dark clouds LDN 306, 314, 315 and 322) that extends $\sim4\arcdeg$ in Sagittarius \citep{Lynds1962, Reipurth2008, Saito1999}. The distances reported in the literature range between 1.5 and 2.4~kpc \citep[e.g.][]{Racine1968, Humphreys1978, Rodriguez1980}.  However, the most used value in the literature is 1.7~kpc. For this distance, the spatial scale of LDN 291 cloud complex  is $\sim$75~pc~$\times$19~pc and the mass is $\sim$1.2~$10^5$~\msun~\citep{Saito1999}.

\subsection{Analysis}\label{App:Ana}
Here we present two different approaches to better estimate the distance to the LDN 291 / GGD 27 region. The two approaches rely on the Gaia DR2 catalogue  \citep{GAIA_2016, GAIA_2018}. First, we used data from Gaia in combination with near-IR polarimetric data of the YSOs located in the GGD 27 region \citep{Kwon16}. 
The second approach is to use the on-line {\it STILISM}\footnote{https://stilism.obspm.fr} application \citep{Capitanio17, Lallement18}, which combines the Gaia data with extinction maps to obtain 3-D dust maps of the Galaxy.  

\subsubsection{Method 1: Young Stellar Objects in GGD 27 and Gaia}
\begin{figure}[!htb]
\begin{center}
\includegraphics[width=0.46\textwidth]{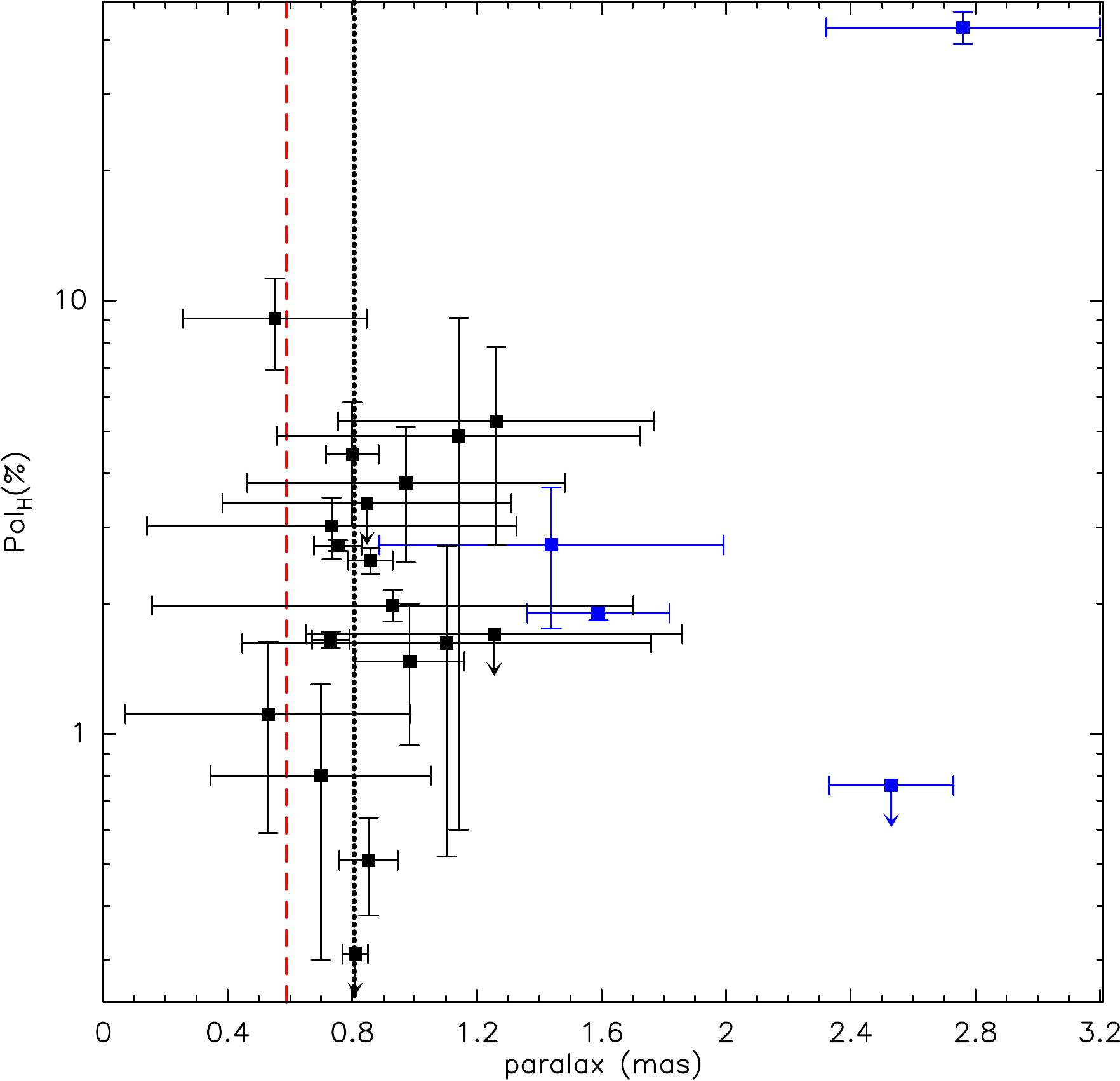}
\caption{
Gaia parallax (in mas) {\it versus} polarization fraction in the $H$ band for the YSO stars that appear in any of the \cite{Kwon16} and \cite{Qiu2009Letter} catalogues. The blue color stars have distances  smaller than $\simeq$1100~pc (considering their error bars), suggesting that they are not associated with the cloud. These stars were not used in the cloud distances estimation.
The black dashed vertical line shows the averaged parallax (weighted by the uncertainty) of the YSOs in GGD~27, 0.81~mas (1.25~kpc). The red dashed vertical line shows the parallax for the distance previously adopted in the literature, 1.7~kpc. 
}
\label{fig:YSO}
\end{center}
\end{figure}

Figure~\ref{fig:YSO} shows the polarization fraction as a function of the Gaia DR2 parallax for the YSOs in GGD~27
(from the \cite{Kwon16} and 
This figure shows that the YSOs population with counterpart in GAIA have distances clearly smaller than the distance adopted in the literature, 1.7~kpc. Most of the YSOs have parallaxes between $\sim$0.6~mas (1.67~kpc) and 1.2~mas (830~pc) but with significant uncertainties. However, there are two objects that have large parallaxes. One of them is [HL85] GGD~27-28 31 \citep{Hartigan85}. It is located at a distance of $362\pm58$~pc but has very high polarization levels \citep[26\%, 42\% and 57\% in the $JHK$ bands, respectively;][]{Kwon16}. This star is located in front of the bright GGD~27 nebula, where high levels of circular polarization are detected. Therefore, the near-IR linear polarization is likely coming from the nebula and it is not related with the optical star.  The other star, 2MASS J18185959-2045537, is located at $395\pm31$~pc and it exhibits no polarization in the near-IR, which indicates that this is a (cold) foreground star not related to the cloud. 
In order to estimate the distance we calculated for the YSOs the average value of the Gaia parallaxes weighted with the uncertainty: 
\begin{equation}
< \pi_{\mathrm{GGD27}} > = \frac{\sum\limits_{i}^{N} \pi_i/\sigma_i^2}{\sum\limits_{i}^{N} 1/\sigma_i^2}
\end{equation}
and the uncertainty is:
\begin{equation}
\sigma(< \pi >) =  \sqrt{\frac{N}{\sum\limits_{i}^{N} 1/\sigma_i^2}}
\end{equation}
$\pi_i$ and  $\sigma_i$ are the parallax for each star and its uncertainty, respectively, from the GAIA DR2 catalogue.
We excluded four stars with parallaxes, within their uncertainties, larger than 0.9~mas (with distances less than $\simeq$1100~pc). Using the rest of the sample, we obtain an average, weighted by the uncertainty, parallax of $0.801\pm0.106$~mas. Therefore, the distance to the YSO cluster is 1248$\pm$166 pc. 

%--------------------------------------------------------
%-------------------------------------------4------------

\subsubsection{Method 2: Extinction-Gaia data -- {\it STILISM}}

\begin{figure}[!h]
\begin{center}
\includegraphics[width=0.45\textwidth]{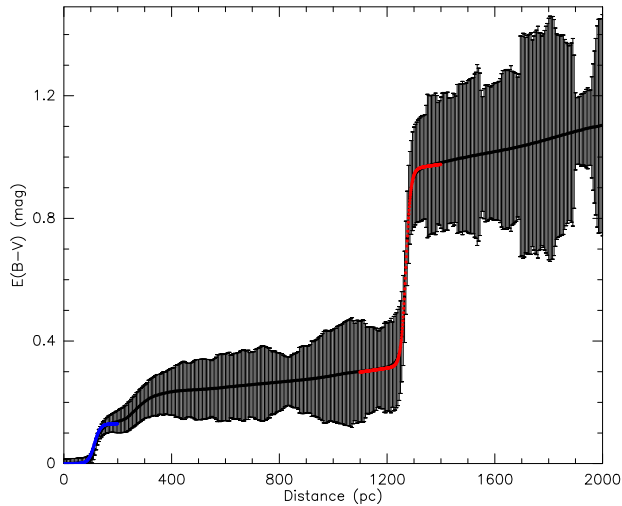}
\caption{
$E(B-V)$ extinction (in magnitudes) as a function of the distance toward the GGD~27  region obtained from the 
{\it STILISM}  (Structuring by Inversion the Local Interstellar Medium)
on-line application \citep{Capitanio17, Lallement18}. The blue and red lines are the fit to the visual extinction jump at 119 and 1270 pc, respectively.
}
\label{fig:EBV}
\end{center}
\end{figure}

A recent work \citep{Danielski18} has correlated the Gaia distances and the corrected version of G extinction with archival ground based data, specially with 2MASS and SDSS/APOGEA-DR14 \citep{Capitanio17, Lallement18}. This allows to derive 3D maps of the extinction as a function of the distance. We used the on-line tool that provides the cumulative reddening curve as a function of the distance for a given line-of-sight.
Figure~\ref{fig:EBV} shows the $E(B-V)$ extinction as a function of the distance toward the LDN~291 / GGD~27 region. It clearly shows two abrupt increases of the extinction, a small one $(E(B-V)\sim$0.2~mag or $A_{\mathrm{V}}$ $\sim$0.6~mag) around 100~pc and a larger one $(E(B-V)\sim$0.7~mag or $A_{\mathrm{V}}$ $\sim$2.0~mag) around 1200~pc. 
In order to fit the two abrupt extinction jumps, we used the following approach:

\begin{equation}
E(B-V) = \frac{A_0}{1+e^{ -A_1(D_{}-D_{\mathrm{jump}}) } }+A_2  D_{}+A_3,
\end{equation}

where $D$ is the distance, $D_{\mathrm{jump}}$ is the distance where the jump occurs and $A_i$ (${\rm i}=0$, ..., 3) are free parameters. We used a reduced $\chi^2$ fit. 
The distance for the large jump is $1270\pm65$~pc. We also used this expression to estimate the first small extinction jump, obtaining a distance of $119\pm15$~pc. 

%--------------------------------------------------------
%-------------------------------------------4------------

\subsubsection{Distance to the GGD~27 nebula/molecular cloud}

The YSOs detected with Gaia are likely near the surface of the cloud, otherwise the cloud extinction would make them not visible at optical wavelengths. Indeed, most of the \textit{Spitzer} YSOs from \citet{Qiu2009Letter} do not have optical counterparts. 
The previous analysis indicates that
the average distance within 1-$\sigma$ uncertainty is between $\sim$1100 to 1400~pc.
The second method is even more very sensitive to the cloud's surface, giving a distance between 1200 and 1340~pc (also at 1 $\sigma$ level). Therefore, combining both methods, we can constrain the distance to the GGD~27 region in the range of 1200 to 1400~pc. 

\section{\textbf{Model robustness} \label{sec:modelrobustness}}

\begin{figure}
    \includegraphics[width=0.47\columnwidth]{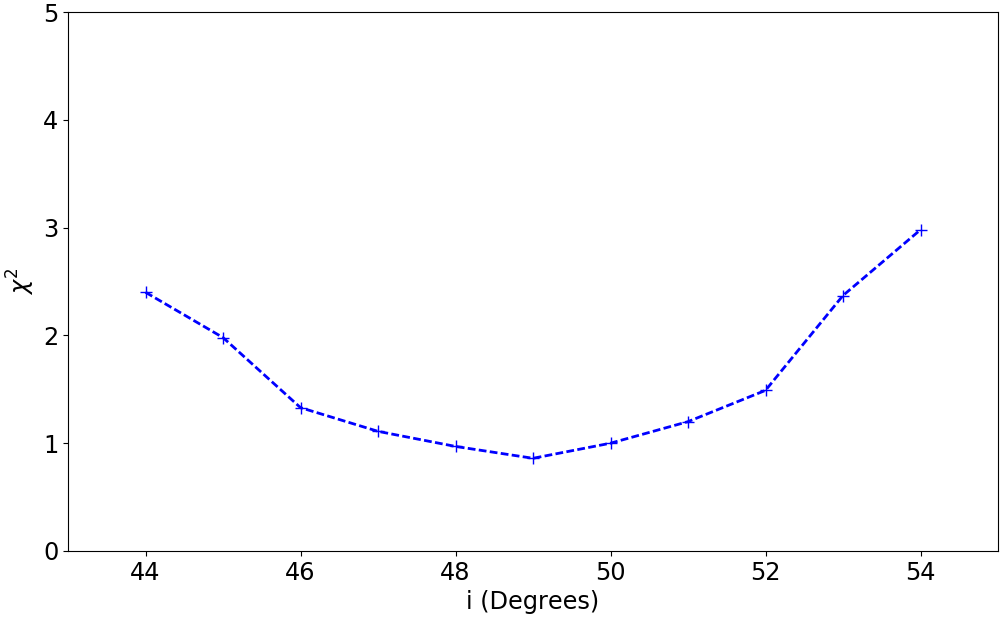}
    \includegraphics[width=0.47\linewidth]{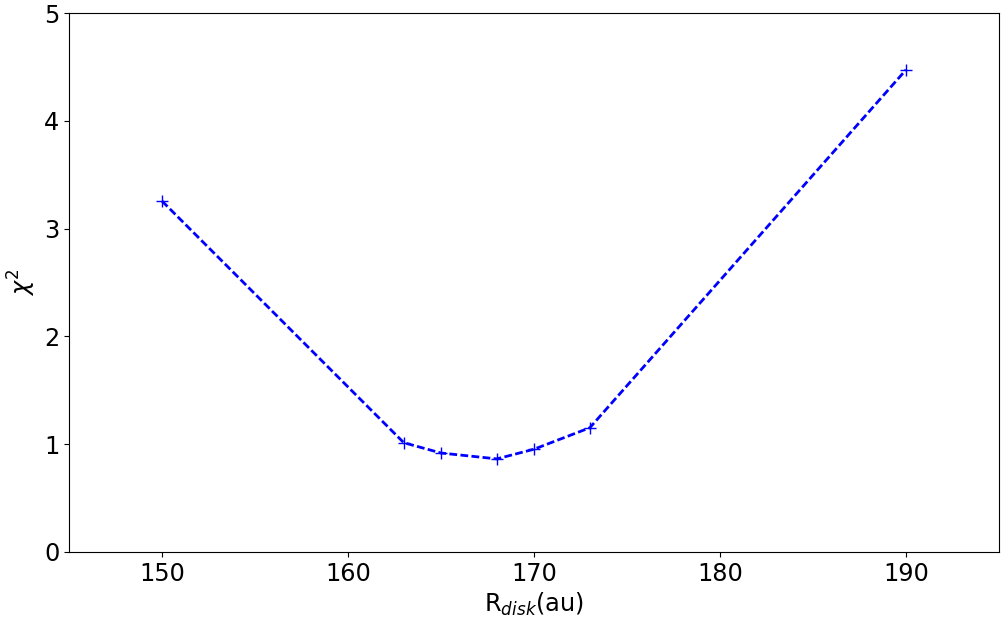}
    
    \includegraphics[width=0.47\linewidth]{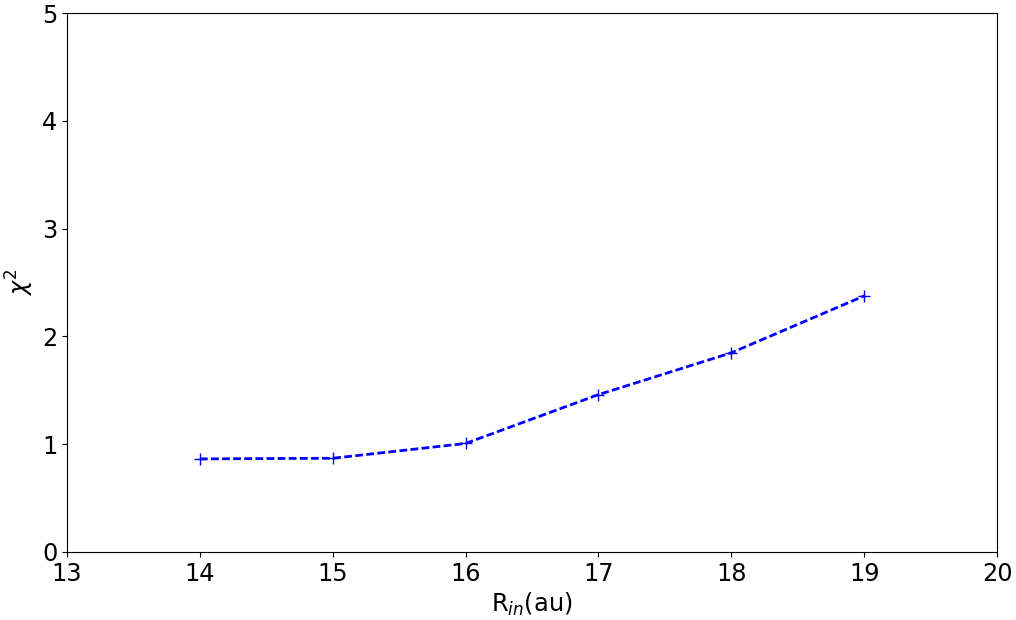}
    \includegraphics[width=0.47\linewidth]{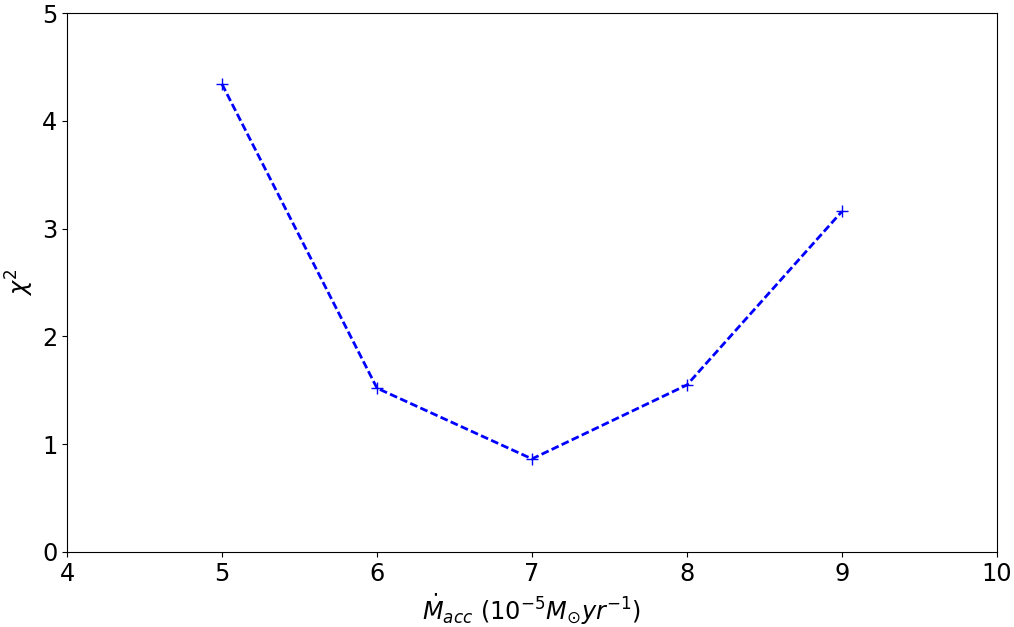}
    
    \caption{\textbf{$\chi^2$ for the best fit model varying inclination (upper left), R$_{\rm disk}$ (upper right), 
    R$_{\rm in}$ (bottom left), \Macc (bottom right). } \label{fig:xi2}}
\end{figure}

As we already advanced in Section \ref{sec:procedure}, Figure \ref{fig:xi2} shows how the $\chi^2$ value changes when we vary the main parameters of the disk, while fixing the others to the best fit value. The same scale for the Y-axis ($\chi^2$) is used in the four panels.
The parameters under study were varied between $\sim$ 10$\%$ and $\sim$ 30$\%$ according to the observational restrictions (see Section \ref{sec:HH80--81}). The four parameters show a minimum at the value of our best fit model. 
The panel shows that the model is much more sensitive to changes in inclination and disk radius, where the parameter variations are around 10$\%$, compared to changes in the
inner radius and accretion rate, where the variations are over 30$\%$.

Regarding the inner radius, we situated the sublimation wall at 12~au ($\pm$ 2~au), based on the luminosity and the dust sublimation temperature. Inward from this border the dust can not survive, thus we can not obtain a physically consistent model with an inner radius smaller than 14~au. In addition, an inner radius larger than $\sim$20~au ($\sim$14~mas) is discarded because it should have been observed with the angular resolution of our ALMA observations.

%%%%%%%%%%%%%%%%%%%%%%%%%%%%%%%%%%%%%%%%%%%%%%%%%%%%%%%%%%%%%%%%%%

\clearpage

\bibliography{hh80}

\end{document}